\documentclass[11pt]{wlscirep}
\usepackage[utf8]{inputenc}

\usepackage{amsmath,amssymb}
\usepackage{type1cm}
\usepackage{mathrsfs}
\usepackage{mathtools}
\usepackage{url}
\usepackage{graphicx}
\usepackage{hyperref}

\usepackage{color}
\usepackage{booktabs}
\usepackage{braket}

\newcommand\HL[1]{{\color{black}#1}}
\newcommand\HLL[1]{{\color{black}#1}}

\usepackage{newtxtext,newtxmath} 

\usepackage{geometry}
\renewcommand{\vec}[1]{\mathbf{#1}}

\title{Demonstration of the rotational viscosity transfer across scales in Navier--Stokes turbulence}

\author[1,*]{Satori Tsuzuki}
\affil[1]{Research Center for Advanced Science and Technology, The University of Tokyo, 4-6-1, Komaba, Meguro-ku, Tokyo 153-8904, Japan}
\affil[*]{tsuzukisatori@g.ecc.u-tokyo.ac.jp}

\begin{abstract}
Mechanical effects that span multiple physical scales---such as the influence of vanishing molecular viscosity on large-scale flow structures under specific conditions---play a critical role in real fluid systems. The spin angular momentum-conserving Navier--Stokes equations offer a theoretical framework for describing such multiscale fluid dynamics by decomposing total angular momentum into bulk and intrinsic spin components. However, this framework still assumes locally non-solid rotational flows, a condition that remains empirically unverified. This study addresses such unvalidated assumptions intrinsic to the model and extends it within the framework of turbulence hierarchy theory. The theory suggests that under certain conditions, small-scale structures may transfer to larger scales through the rotational viscosity. To verify this, we conducted spectral analyses of freely decaying two-dimensional turbulence initialized with a vortex-concentrated distribution. The results indicate that rotational viscosity exhibits interscale transfer behavior, revealing a new mechanism by which order can propagate from small to large scales in Navier--Stokes turbulence.
\end{abstract}
\begin{document}

\flushbottom
\maketitle
%
%
\thispagestyle{empty}
\section{Introduction}
Turbulence exhibits intricate hierarchical and ordered structures, posing a longstanding challenge in fluid dynamics. A fundamental understanding of these features has been pursued from both statistical-mechanics and dynamics perspectives. Milestones in the statistical-mechanics approach include Onsager's prediction of inverse energy cascades in two-dimensional (2D) turbulence and Kraichnan's spectral theory, both of which provided insights into energy transfer mechanisms~\cite{Onsager1949, Kraichnan1967, Frisch1995, Davidson2015}. Multiscale dynamics---including microscopic phenomena influencing macroscopic flow structures---play a critical role in real fluid systems. A prominent example is found in macroscopic quantum fluids~\cite{Donnelly1991, Barenghi2014}, where the near-zero molecular viscosity at cryogenic temperatures enables the formation of quantum vortices with minimal dissipation. These quantum-scale features can, in turn, affect large-scale dynamics, giving rise to striking behaviors such as the fountain effect and film flow. Such phenomena underscore the need for a multiscale fluid dynamical model that can consistently account for interactions across disparate physical scales. One promising candidate is the spin-angular-momentum-conserving Navier--Stokes model, originally proposed in 1964~\cite{doi:10.1063/1.1711295, Eringen1966}. This extended formulation provides a framework for describing multiscale fluid dynamics bridging classical and quantum regimes. The model distinctly separates the local total angular momentum into two components: the contribution from the bulk fluid and the internal spin degrees of freedom associated with the constituent particles, such as molecules. This decomposition permits the inclusion of internal rotational motion---specifically, particle-scale angular velocity---into the continuum-level equations of motion. While mathematically consistent with the conventional Navier--Stokes equations, this formulation is particularly well-suited for systems in which internal rotation affects macroscopic behavior. Applications include polar fluids~\cite{10.1122/1.3302803}, complex suspensions~\cite{10.1063/5.0087981}, and other structured media where microscopic rotation is non-negligible. More recently, it has been suggested that this model possesses a mathematical structure capable of mediating inter-scale coupling, allowing microscale interactions (such as among small-scale vortices) to influence macroscale flow features, as discussed in section 3.2 of Ref.~\cite{tsuzuki2025multi}. This potential has renewed interest in the model as a candidate framework for multiscale fluid dynamics. 

However, this formulation assumes that local rotational motion can be non-solid, meaning the spinning of constituent fluid particles does not always coincide with solid body rotation. Specifically, when the local fluid velocity is denoted by $\vec{u}$, the vorticity by $\boldsymbol{\omega}$, and the angular velocity by $\boldsymbol{\omega}_{0}$, the model assumes that $\boldsymbol{\omega}$ is not always equal to $2\boldsymbol{\omega}_{0}$. This assumption is uncommon in classical fluid mechanics because the local nature of fluid particles usually ensures solid rotational flow at every point. Verifying this assumption is challenging, as it requires understanding the relationship between vortex stretching and tilting geometric features~\cite{KAWAHARA_KIDA_TANAKA_YANASE_1997} and the deformation of fluid particles. Furthermore, in 2D problems where vortex stretching and tilting are absent, a theoretical justification of the assumption is necessary before the model can be confidently applied.

This study addresses this open question by developing a continuum-mechanical derivation of the $\boldsymbol{\omega} = \nabla \times \vec{u} \ne 2 \boldsymbol{\omega}_{0}$ condition in 2D, and by exploring its validity through numerical simulations.
Specifically, we examine a freely decaying 2D turbulent flow initialized with a set of point vortices exhibiting localized circulation. This configuration mimics the state of a system immediately following the release of externally controlled condition---for example, in phase-modulated polar or quantum fluids. Our analysis shows that the quantity $\boldsymbol{\omega} - 2\boldsymbol{\omega}_{0}$ is directly proportional to the first-order spatial displacement, making it non-negligible under specific conditions. In addition, the spin angular momentum conserving Navier--Stokes equations, when extended to a multiscale physics framework, theoretically suggest that rotational viscosity \HLL{$\nabla \times \boldsymbol{\omega}$} can be transferred to larger scales under particular conditions. This study investigates the particular conditions under which local non-solid rotational flow occurs and proposes that rotational viscosity can be transferred between scales in a system with a concentrated point vortex and circulation. To verify this, spectral analyses were conducted on freely decaying 2D turbulence that was initialized with a distribution of concentrated vortices. The results show that rotational viscosity exhibits interscale transfer behavior, revealing a novel mechanism by which order can propagate from small to large scales in Navier--Stokes turbulence.

\HLL{The remainder of this paper is organized as follows. In Sec.~\ref{sec:GEOFMultiFluidMech}, we first review the representative derivation of the spin angular momentum-conserving Navier--Stokes equations by D.~W.~Condiff, and highlight two central issues: (i) the \textit{a priori} assumption of locally non-solid rotational flow, and (ii) the need to introduce equations of motion for internal degrees of freedom based on a molecular interpretation of fluid elements. In the latter part of the same section, we summarize our recent work showing that this molecular interpretation can be avoided by adopting a multiscale definition of vorticity (later expressed in Eq.~\ref{eq:TransOmega}), and that the governing equations can be closed by modeling the influence of residual quantum vortices that persist below the Kolmogorov scale due to extremely low viscosity and negligible thermal dissipation.

In Sec.~\ref{sec:eringen}, we rederive the spin-conserving Navier--Stokes equations using Eringen's micropolar fluid theory~\cite{ERINGEN1964205, ERINGEN1966JMM, ERINGEN1969115}, one of the most established frameworks in this area. Unlike Condiff's approach, Eringen's formulation does not rely on molecular or particulate assumptions, offering a more consistent continuum-mechanical interpretation. Nevertheless, it shares the same core issues---namely, the assumption of locally non-solid rotation and the emergence of additional variables related to internal angular motion (e.g., microrotation or couple stresses). In both formulations, a multiscale treatment such as Eq.~\ref{eq:TransOmega}, which maps these internal variables onto subscale quantities, becomes essential.

In Sec.~\ref{sec:rederivmulti}, we show that Eq.~\ref{eq:TransOmega} can be interpreted as the Green's function solution to a Poisson-type equation with subgrid-scale (SGS) stress as the source term. This arises when the vorticity equation is filtered under the large-eddy simulation (LES) framework, assuming incompressibility and stationarity. We argue that this provides a physically grounded and computationally practical closure model for the SGS stress, based on local turbulent scaling. We also discuss its mathematical properties, including stability and convergence over time.

This theoretical development suggests that the SGS stress can be expressed in terms of the velocity field, vorticity, and higher-order quantities such as $\nabla \times \boldsymbol{\omega}$. This highlights the role of $\nabla \times \boldsymbol{\omega}$ in mediating small-scale dynamics within the Navier--Stokes system---a key implication of incorporating spin angular momentum conservation.
In Sec.~\ref{seq:rethinknonsolid}, we examine the second foundational requirement for this formulation: the physical presence of locally non-solid rotational flow.

In Sec.~\ref{sec:simulation}, we validate these theoretical insights through numerical simulations of two-dimensional freely decaying turbulence, initialized with a point vortex distribution. We conduct spectral analyses of $\nabla \times \boldsymbol{\omega}$ and identify regimes where rotational viscosity effects become significant. Finally, Sec.~\ref{sec:conclusion} summarizes our findings and discusses potential directions for future work.}

\section{Governing equations of multiscale fluid mechanics} \label{sec:GEOFMultiFluidMech}
{\it Model~overview.}---The Navier--Stokes equation incorporating spin angular momentum conservation was derived by D.W. Condiff in 1964 through the following procedure. First, the local total angular momentum $\vec{M}$ is decomposed into two parts: the angular momentum of the bulk fluid $(\vec{r} \times \vec{u})$ and the internal degrees of freedom of the constituent particles $(\vec{l})$ as $\vec{M} = \vec{r} \times \vec{u} + \vec{I}$. The vector $\vec{I}$ is the spin angular momentum intrinsic to each constituent particle. Vectors $\vec{r}$ and $\vec{u}$ represent the coordinates and velocity of the local fluid fragment, respectively. Vector $\vec{I}$ can be further expressed as the product of the moment of inertia of the constituent particles and the spin field defined on the constituent particles as $\vec{I} = \bar{\vec{I}}\cdot \boldsymbol{\omega}_{0}$ where $\bar{\vec{I}}$ is the tensor field, expressed as a scalar multiple of the unit dyadic $\vec{U}$ as $\bar{\vec{I}} = {\it I}\vec{U}$ and thereby $\bar{\vec{I}}\cdot \boldsymbol{\omega}_{0} = {\it I}\vec{U}\boldsymbol{\omega}_{0} = {\it I}\boldsymbol{\omega}_{0}$, where $\boldsymbol{\omega}_{0}$ is now represented as a vector. This relation assumes a uniform and isotropic spin field for each constituent particle. In this case, $\boldsymbol{\omega}_{0}$ represents the angular velocity vector around the axis of the constituent particle. 

Substituting these relations ($\vec{M} = \vec{r} \times \vec{u} + \vec{I}$ and $\vec{I} = {\it I}\boldsymbol{\omega}_{0}$) and the following assumptions into the three equations: a) Cauchy's equation of motion, b) the conservation law for the local angular momentum $\vec{M}$ derived from Reynolds' transport theorem, and c) the evolution equation for the spin angular momentum $\vec{I}$ derived from these equations, we obtain the spin-angular-momentum conserving Navier--Stokes equation. The two assumptions mentioned above are as follows: (1) Newtonian fluid assumption---This allows us to divide the stress tensor $\vec{T}$ into a symmetric part $\vec{T}_{s}$ and an asymmetric part $\vec{T}_{a}$ of the stress tensor, where $\vec{T}_{s}$ is proportional to the symmetrized velocity gradient tensor. The couple stress tensor is assumed to depend only on the symmetrized spin gradient tensor.
(2) Possibility of non-solid rotation of fluid particles---If the local vorticity is $\boldsymbol{\omega} = \nabla \times \vec{u}$ and the angular velocity of the constituent particles at that location is $\boldsymbol{\omega}_{0}$ and the deviation from the solid rotating flow contributes to the asymmetric part of the stress tensor: $\vec{T}_{a} = \eta_{r} (\boldsymbol{\omega} - 2\boldsymbol{\omega}_{0}$). The resulting hydrodynamic equation is described as follows: 
\begin{equation}
\frac{{D} \vec{u}}{{D} t} = -\frac{1}{\rho}\nabla P + (\eta + \eta_{r})\nabla^2 \vec{u} 
+ \bigl( \frac{\eta}{3} + \xi -\eta_{r} \bigr) \nabla\nabla\cdot\vec{u} 
+ 2\eta_{r}\nabla\times\boldsymbol{\omega}_{0} + \vec{F}, \label{eq:NSwithSAMC}
\end{equation}
where $D \{ \cdot \}/D t$ is the material derivative. $P$ is pressure. $\vec{u}$ denotes the velocity of a constituent particle, and $\boldsymbol{\omega}_{0}$ represents the angular velocity of the particle around its axis. $\vec{F}$ is an external force. The parameter $\eta$ signifies the shear viscosity, $\xi$ indicates the bulk viscosity, and $\eta_{r}$ represents the rotational viscosity coefficients, respectively.  A detailed step-by-step derivation of Eq.~(\ref{eq:NSwithSAMC}) is presented in Fig. 9 of Ref.~\cite{10.1063/5.0218444} in a tabular format.

The resulting Eq.~(\ref{eq:NSwithSAMC}) represents an extension of the conventional Navier--Stokes equation. When the rotational viscosity $\eta_{r}$ is set close to zero, the fourth term on the right-hand side vanishes. In this case, only the shear viscosity in the first term and the shear and bulk viscosities in the second term on the right-hand side remain. As a result, Eq.~(\ref{eq:NSwithSAMC}) reduces to the usual compressible Navier--Stokes equations. Since this study focuses only on incompressible fluids, the incompressibility condition $\nabla \cdot \vec{u} = 0$ always holds, implying that the third term on the right-hand side becomes zero. Therefore, the only distinctive feature of Eq.~(\ref{eq:NSwithSAMC}) compared to the usual Navier--Stokes equation is the rotation term in the fourth term on the right-hand side.

The angular velocity $\boldsymbol{\omega}_0$ was originally introduced not as a characteristic of fluid constituents (i.e., fluid particles), but as a molecular-scale variable. Specifically, the local angular momentum $\vec{M}$ is decomposed into the angular momentum of the bulk flow, $\vec{r} \times \vec{u}$, and the internal degrees of freedom (e.g., spin) of individual molecules. However, this interpretation is inconsistent with classical fluid mechanics, wherein molecules are regarded as infinitesimal, and no deformation is assumed at a point \HLL{mass}. Consequently, molecular-level rotation is treated such that $\boldsymbol{\omega} = 2\boldsymbol{\omega}_0$. \HLL{In contrast}, interpreting $\boldsymbol{\omega}_0$ as a fluid particle variable also contradicts classical fluid mechanics in a strict sense. Fluid particles (or fluid percels) are typically defined as small virtual element with a characteristic size that is smaller than the Kolmogorov scale, which is the minimum scale of an eddy, but is much larger than the free paths of molecules~\cite{batchelor2000introduction, bennett2006lagrangian, HAERI2011716}. Under this assumption, local rotational flow within a fluid particle is considered solid-body rotation, implying that $\boldsymbol{\omega} = 2\boldsymbol{\omega}_0$ always holds. As a result, $\vec{T}_{a} = \eta_{r} (\boldsymbol{\omega} - 2\boldsymbol{\omega}_{0}$) vanishes, rendering Eq.~(\ref{eq:NSwithSAMC}) physically irrelevant. The case of compressible fluids is no exception---while bulk deformation may occur, deformation of fluid particles, being sufficiently small compared to the macroscopic length scale, is rarely considered.
However, in mesoscale dynamics such as molecular dynamics or dissipative particle dynamics (DPD)~\cite{PJHoogerbrugge_1992, JMVAKoelman_1993, MULLER2015301}, fluid particles are interpreted as coarse-grained entities~\cite{PhysRevE.57.2930, doi:10.1142/S0129183197000771, DZWINEL20062169} representing groups of atoms or molecules, which can undergo deformation. In such frameworks, local vortex stretching and tilting can be defined at the level of fluid particles, and the deviation between $\boldsymbol{\omega}$ and $2\boldsymbol{\omega}_0$ naturally contributes to the antisymmetric part of the stress tensor. Under these conditions, Eq.~(\ref{eq:NSwithSAMC}) becomes valid. This empirical justification supports the use of $\boldsymbol{\omega}_0$ as a fluid-particle variable in mesoscale dynamics.

Nevertheless, interpreting $\boldsymbol{\omega}_0$ as a variable defined on coarse-grained particles that represent atoms or molecules introduces a challenge: establishing a relationship between $\boldsymbol{\omega}_0$ and the actual microscopic rotational behavior of molecules. In other words, the material properties associated with $\boldsymbol{\omega}_0$ must be well defined. Establishing a quantitative relationship between $\boldsymbol{\omega}$---a continuum scale variable---and the microscopic parameters of atoms and molecules is far from trivial. Although the derivation of Eq.~(\ref{eq:NSwithSAMC}) is accompanied by a time-evolution equation for variable $\boldsymbol{\omega}_0$ (see Eq.~(13) in Ref.~\cite{doi:10.1063/1.1711295} if needed), this equation cannot be solved deterministically for several real cases due to the lack of an initial condition for $\boldsymbol{\omega}_0$. Put simply, the system of equations is underdetermined by one equation governing $\boldsymbol{\omega}_0$. For these reasons, Eq.~(\ref{eq:NSwithSAMC}) has been largely limited to applications in chemical engineering and related fields.

{\it Spin-field correspondence.}---A critical issue is that $\boldsymbol{\omega}_{0}$ is defined only ``on'' individual particles and not as a field variable. This leads to a problem when applying the relation $\vec{T}_{a} = \eta_{r}(\boldsymbol{\omega} - 2\boldsymbol{\omega}_{0})$ at the particle scale, as it becomes meaningless if $\boldsymbol{\omega} = 2\boldsymbol{\omega}_{0}$. Alternatively, if the relation holds over a larger region, the continuity condition for $\boldsymbol{\omega}_{0}$ is not preserved at the particle scale. 
A recent proposal in Ref.~\cite{tsuzuki2025multi} offers a viable solution to this inconsistency. The underlying concept involves substituting the spin variable $\boldsymbol{\omega}_{0}$ with a field variable $\overline{\boldsymbol{\omega}}$. This approach is grounded in the notion that the fundamental fluid unit can be characterized as a field, rather than a particle. This perspective shares similarities with the turbulence hierarchy, which effectively decomposes turbulence into distinct eddy structures. A constitutive relationship is then established between $\overline{\boldsymbol{\omega}}$ and $\boldsymbol{\omega}_{0}$, thereby reconciling the local and nonlocal viewpoints.
\begin{align}
\HL{\overline{\boldsymbol{\omega}}(\vec{r}) \coloneqq \frac{1}{C} \int \frac{\boldsymbol{\omega}_{0}(\vec{r}')}{|\vec{r} - \vec{r}'|}\, d\vec{r}',} \label{eq:TransOmega} 
\end{align}
where $C$ is a constant coefficient. 
The transformation between $\boldsymbol{\omega}_{0}$ and $\overline{\boldsymbol{\omega}}$ as defined by Eq.~(\ref{eq:TransOmega}) can be expressed as $\overline{\boldsymbol{\omega}}$ = ${\mathcal F}[\boldsymbol{\omega}_{0}]$, where ${\mathcal F}[\cdot]$ denotes the convolution integral operator on the right-hand side of Eq.~(\ref{eq:TransOmega}). The internal inconsistency in Eq.~(\ref{eq:NSwithSAMC}), which requires $\boldsymbol{\omega} = 2\boldsymbol{\omega}_{0}$ at each fluid particle even if $\boldsymbol{\omega} \ne 2\boldsymbol{\omega}_{0}$, can be resolved by reinterpreting Eq.~(\ref{eq:TransOmega}) within the framework of hierarchical concept of turbulence theory. Specifically, Eq.~(\ref{eq:TransOmega}) may be understood as defining a \HLL{mapping} from \HLL{the $\boldsymbol{\omega}_{0}$-based vorticity field, which encapsulates small-scale information, to the coarse-grained vorticity} $\overline{\boldsymbol{\omega}}$, mediated by the operator ${\mathcal F}[\cdot]$. 
From a large-scale perspective, the motion of a small-scale vortex is sufficiently localized to be considered as localized solid-body rotation. Consequently, we acknowledge that the difference between $\boldsymbol{\omega}$ and $2\boldsymbol{\omega}_{0}$ may be nonzero when examining the small-scale vortices from the same level, while the small scale vorticity can still be approximated as $\boldsymbol{\omega} = 2\boldsymbol{\omega}_{0}$ when viewed from a large-scale perspective. Therefore, $\boldsymbol{\omega}_{0}$ on the right-hand side can be considered as $(1/2)\boldsymbol{\omega}$, as acknowledged in classical fluid mechanics, and this relation holds across scales. Then, vorticity $\vec{\overline{\omega}}$ at a large-scale is obtained from the convolution integral of the vortices characterized by $\boldsymbol{\omega}~(=2\boldsymbol{\omega}_{0})$ at its subscale. Notably, Eq.~(\ref{eq:TransOmega}) only states that the vorticity at a given scale is obtained from the vorticity distribution at a finer subscale, without implying any discontinuity in the fluid; no singularities exist at position $\vec{r}$.

The relationship between $\boldsymbol{\omega}_{0}$ and $\overline{\boldsymbol{\omega}}$ is analogous to the relationship in electromagnetism between the steady current density $\vec{i}$, the resulting magnetic field $\vec{B}$, and the vector potential $\vec{A}$ satisfying $\vec{B} = \nabla \times \vec{A}$. In fact, $\overline{\boldsymbol{\omega}}$ plays the role of a rotational potential induced by the spatial distribution of the local angular velocity $\boldsymbol{\omega}_{0}$ of fluid particles. The curl $\nabla \times \overline{\boldsymbol{\omega}}$ thus represents the rotational force induced at position $\vec{r}$, which can be interpreted as a form of viscosity. Specifically, $\nabla \times \boldsymbol{\omega} = \nabla \times (\nabla \times \vec{u}) = \nabla(\nabla \cdot \vec{u}) - \nabla^2 \vec{u} = -\nabla^2 \vec{u}$, where we have used the incompressibility condition $\nabla \cdot \vec{u} = 0$. Hence, $\nabla \times \boldsymbol{\omega}$ effectively conveys the rotational viscosity.

Replacing $\boldsymbol{\omega}_{0}$ with $\overline{\boldsymbol{\omega}}$ in the derivation of Eq.~(\ref{eq:NSwithSAMC}) and substituting $(1/2)\boldsymbol{\omega}$ into Eq.~(\ref{eq:TransOmega}), the rotational contribution in Eq.~(\ref{eq:NSwithSAMC}) can be further reformulated as follows using simple vector relations: 
\begin{equation}
\nabla \times \overline{\boldsymbol{\omega}}(\vec{r}) ~=~ \biggr ( \frac{1}{2C} \biggl ) \int \frac{\boldsymbol{\omega}(\vec{r}') \times (\vec{r} - \vec{r}')}{|\vec{r} - \vec{r}'|^{3}} \, d\vec{r}' ~+~ {\mathcal F} \Bigl [\nabla \times \boldsymbol{\omega}(\vec{r})\Bigr ]. 
\label{eq:RecurrRel}
\end{equation}
Denote the hierarchical depth level of vortex structures by an index $n$, such that large-scale vortices are associated with depth $n$, and subgrid-scale vortices correspond to depth $n + 1$. 
In Eq.~(\ref{eq:RecurrRel}), the left-hand side represents the vorticity at depth $n$, while the first term on the right-hand side represents vortex--vortex interactions at depth $n + 1$. The second term on the right-hand side corresponds to the rotational field $\nabla \times \boldsymbol{\omega}$ at depth $n + 1$, filtered through the operator ${\mathcal F}$. Notably, Eq.~(\ref{eq:RecurrRel}) can be expressed as a recursive relation of the form $\vec{X}^{(n)} = \vec{Y}^{(n+1)} + {\mathcal F}[\vec{X}^{(n+1)}]$, where $\vec{X} = \nabla \times \boldsymbol{\omega}$, and $\vec{Y}$ denotes the vortex interaction term at depth $n + 1$. Within this hierarchical framework, a generalized form of Eq.~(\ref{eq:RecurrRel}) from depth $n$ to $k$ can be written as

\begin{eqnarray}
\mathbf{X}^{(n)} = \mathbf{Y}^{(n+1)} + \mathcal{F} \Big( 
  \mathbf{Y}^{(n+2)} + \mathcal{F} \Big( 
    \mathbf{Y}^{(n+3)} + \cdots + \mathcal{F} \left[ \mathbf{X}^{(n+k)} \right] 
  \Big) 
\Big). \label{eq:GeneralRecurrRel}
\end{eqnarray}

Let us now consider a case in which the subgrid scale reaches the Kolmogorov minimum scale, but the vortex remains stable. A representative example is the quantum vortex in liquid helium-4 cooled to cryogenic temperatures. Since no smaller vortices exist, the second term on the right-hand side of Eq.~(\ref{eq:RecurrRel}) is zero. On the other hand, the vorticity $\boldsymbol{\omega}$ is localized on the vortex filaments and thus is expressed as
\begin{eqnarray}
\HL{\boldsymbol{\omega}(\vec{r})~=~\kappa \int_{\varGamma} \vec{s}'(\chi) \delta (\vec{r} - \vec{s}(\chi)) \, d\chi}, \label{eq:CondensVort}
\end{eqnarray}
where $\chi$ is the arc length along the vortex filaments, and $s$ is the position vector from the origin to point $\chi$ on the filaments~\cite{TSUBOTA2013191, Galantucci2020}. The line integral $\varGamma$ is taken along the vortex filaments, and $\kappa$ denotes the intensity of circulation. The vectors $\vec{s}'$ and $\vec{s}''$ are defined such that $\vec{s}'$ is the tangent vector at point $\chi$, and $\vec{s}''$ is the vector perpendicular to $\vec{s}'$. Additionally, $\vec{s}'$, $\vec{s}''$, and the cross product ($\vec{s}' \times \vec{s}''$) are mutually orthogonal. The partial derivatives of $\vec{s}$ with respect to $\chi$ are given by $\frac{\partial \vec{s}}{\partial \chi} = \vec{s}'$. After applying the condition that the second term on the right-hand side of  Eq.~(\ref{eq:RecurrRel}) is zero, we substitute Eq.~(\ref{eq:CondensVort}) into Eq.~(\ref{eq:RecurrRel}) to obtain the relationship between the smallest-scale vortex and the vortex at the next scale.

\begin{eqnarray}
\HL{\nabla \times \overline{\boldsymbol{\omega}}(\vec{r})
~=~\biggr ( \frac{\kappa}{2C} \biggl ) \int_{\varGamma} \frac{\vec{s}'(\chi) \times (\vec{r} - \vec{s}(\chi))}{|\vec{r} - \vec{s}(\chi)|^{3}} \, d \chi.} \label{eq:BiotSavFin}
\end{eqnarray}
The right-hand side of Eq.~(\ref{eq:BiotSavFin}) represents vortex-vortex interactions within a point vortex system comprising multiple vortices. This corresponds to a special case in which vortices do not dissipate at the smallest scale. The rotational forcing at this minimal scale, as expressed on the right-hand side of Eq.~(\ref{eq:BiotSavFin}), is transmitted to larger scales through a recursive nesting structure---mathematically, a composition of recursive transformations---as described in Eqs.~(\ref{eq:RecurrRel}) or (\ref{eq:GeneralRecurrRel}), ultimately reaching the large-scale regime via the rotational term in Eq.~(\ref{eq:NSwithSAMC}). In summary, by redefining the spin variable, originally defined only on discrete particles (fluid elements or molecules) in Eq.~(\ref{eq:NSwithSAMC}), as a continuous field variable through Eq.~(\ref{eq:TransOmega}), we have demonstrated that the spin-angular-momentum-conserving Navier--Stokes equation inherently includes a mechanism for transferring vortex dynamics from small to large scales.
In practical systems with significant dissipation, the dissipation effects of the scale transformation governed by the filtering function ${\mathcal F}$ alone may not be sufficient. Nonetheless, the underlying principle remains valid. In such cases, it is sufficient to reinterpret Eq.~(\ref{eq:GeneralRecurrRel}) as incorporating a composite filtering operation, ${\mathcal F} \circ {\mathcal G}$, where another filtering function ${\mathcal G}$ accounts for enhanced dissipation or scale-localized attenuation.

\section{\HLL{Recovering the spin-conserving Navier--Stokes equation from micropolar fluid theory}} \label{sec:eringen}
\HLL{The most distinctive feature of Eq.~(\ref{eq:NSwithSAMC}) lies in the rotational viscosity term appearing as the fourth term on the right-hand side. Similar local spin angular velocity terms arise in various theories of functional fluids. In the case of magnetic fluids (ferrofluids), the molecular spins are aligned by an external magnetic field, coupling the magnetic moment with the flow field. The fluid theories of Rosensweig~\cite{ROSENSWEIG01051988, rosensweig2013ferrohydrodynamics} and Shliomis~\cite{shliomis1972effective, 10.1063/1.868108} for ferrofluids incorporate rotational equations of motion analogous to Eq.~(\ref{eq:NSwithSAMC}) into the Navier--Stokes framework. Likewise, in the Ericksen--Leslie equations~\cite{Ericksen1962, Leslie1968} for nematic liquid crystals, rotational dynamics and symmetric/antisymmetric stresses stemming from Frank elasticity and Leslie viscosity contribute to the molecular orientation dynamics. Furthermore, in the field of active matter, fluids composed of active particles possessing intrinsic spin---such as magnetically driven systems, bacterial rotors, and chiral active particles---exhibit active torques that couple to the macroscopic rotational field $\boldsymbol{\omega}$, leading to the emergence of rotational terms in the Navier--Stokes equations~\cite{PhysRevLett.106.218101, Ramaswamy_2017, Maitra2019, annurev-conmatphys-040821-125506}. Notably, Eringen's micropolar theory~\cite{Eringen1966} serves as the prototypical mathematical framework underlying all of these descriptions, offering a classical continuum theory that accounts for internal degrees of freedom (spin) of particles. In this section, we theoretically derive Condiff's spin angular momentum-conserving Navier--Stokes equations from Eringen's micropolar theory, thereby demonstrating that they correspond to a special case of the micropolar framework.}

\HLL{
According to Eringen's micropolar fluid theory, each material point possesses a translational velocity field $\mathbf{u}(\mathbf{r},t)$ and an independent micro-rotational angular velocity $\boldsymbol{\omega}_0(\mathbf{r},t)$. The intrinsic angular momentum per unit mass is given by $J \boldsymbol{\omega}_0$, where $J$ is the microinertia. Accordingly, the total angular momentum per unit volume is given by
\begin{equation}
\rho\vec{M} = \rho \mathbf{r} \times \mathbf{u} + \rho J \boldsymbol{\omega}_{0}.
\label{eq:total-angular-momentum}
\end{equation}
The local form of the angular momentum balance is obtained via the divergence theorem applied to a control volume, yielding
\begin{equation}
\rho \frac{D \vec{M}}{Dt} 
= \mathbf{r} \times \rho \mathbf{f} + \rho \mathbf{c} + \nabla \cdot (\mathbf{r} \times \mathbf{T} + \mathbf{m}),
\label{eq:angular-momentum-balance}
\end{equation}
where $\mathbf{c}$ is the external couple per unit mass and $\mathbf{m}$ is the couple stress tensor.
$\mathbf{m}$ is generally modeled as a linear function of the micro-rotation gradients, given by
\begin{equation}
\mathbf{m} = \alpha_1 \nabla \boldsymbol{\omega}_{0} 
+ \alpha_2 (\nabla \boldsymbol{\omega}_{0})^\top 
+ \alpha_3 (\nabla \cdot \boldsymbol{\omega}_{0})\mathbf{I}.
\label{eq:mmeaning}
\end{equation}
Here, each term on the right-hand side can be summarized as follows: 
\begin{itemize}
    \item The term $\alpha_1 \nabla \boldsymbol{\omega}_{0}$ captures the direct coupling between the gradient of the micro-rotation and the couple stress.
    \item The term $\alpha_2 (\nabla \boldsymbol{\omega}_{0})^\top$ represents effects due to the transposed gradient, accounting for possible asymmetric responses.
    \item The term $\alpha_3 (\nabla \cdot \boldsymbol{\omega}_{0})\mathbf{I}$ describes isotropic dilatational contributions due to the divergence of the micro-rotation field.
\end{itemize}
Equation~\eqref{eq:mmeaning} represents a general constitutive model that allows for anisotropic and more complex couplings between micro-rotations and the stress state.
For the purpose of deriving Condiff's model, we will later consider a simplified isotropic case.

The balance of linear momentum (translational motion) takes the form
\begin{equation}
\rho \frac{D\mathbf{u}}{Dt} = \nabla \cdot \mathbf{T} + \rho \mathbf{f},
\label{eq:translational-balance}
\end{equation}
where $\mathbf{T}$ is the generally non-symmetric stress tensor and $\mathbf{f}$ is the external body force.
Taking the cross product of Eq.~\eqref{eq:translational-balance} yields
\begin{equation}
\mathbf{r} \times \left(\rho \frac{D\mathbf{u}}{Dt}\right)
= \mathbf{r} \times (\nabla \cdot \mathbf{T}) + \mathbf{r} \times \rho \mathbf{f}.
\label{eq:translational-cross}
\end{equation}
Recognizing that 
\[
\rho \frac{D}{Dt}(\mathbf{r} \times \mathbf{u}) = \mathbf{r} \times \left(\rho \frac{D\mathbf{u}}{Dt}\right),
\]
we subtract Eq.~\eqref{eq:translational-cross} from Eq.~\eqref{eq:angular-momentum-balance}, yielding
\begin{equation}
\rho J \frac{D\boldsymbol{\omega}_{0}}{Dt}
= \nabla \cdot (\mathbf{r} \times \mathbf{T} + \mathbf{m})
- \mathbf{r} \times (\nabla \cdot \mathbf{T})
+ \rho \mathbf{c}.
\label{eq:spin-evolution-raw}
\end{equation}

Next, we decompose the stress tensor into symmetric and antisymmetric parts
\begin{equation}
\mathbf{T} = \mathbf{T}_{s} + \mathbf{T}_{a},
\label{eq:stress-decomposition}
\end{equation}
with
\begin{equation}
\mathbf{T}_{s} = -P\mathbf{I} + \mu [\nabla \mathbf{u} + (\nabla \mathbf{u})^\top] + \lambda (\nabla \cdot \mathbf{u})\mathbf{I}, \label{eq:symmetric-stress} 
\end{equation}
and
\begin{equation}
\mathbf{T}_{a} = \mu_r \bigl[ (\nabla \mathbf{u}) - (\nabla \mathbf{u})^\top - 2\mathbf{W} \bigr]. \label{eq:antisymmetric-stress}
\end{equation}
Here, $\mu$, $\lambda$, and $\mu_r$ denote the shear viscosity, bulk viscosity, and rotational viscosity coefficients, respectively.
$\mathbf{W}$ is the skew-symmetric tensor associated with $\boldsymbol{\omega}_{0}$:
\begin{equation}
\mathbf{W}_{ij} = -\epsilon_{ijk} \omega^{0}_{k},
\label{eq:W-definition}
\end{equation}
where the lower subscript 0 is shifted to the upper subscript in component expression.
Using standard vector-tensor identities, we find
\begin{equation}
\nabla \cdot (\mathbf{r} \times \vec{T}_{s}) - \mathbf{r} \times (\nabla \cdot \vec{T}_{s}) = 0,
\label{eq:T-symmetric-express}
\end{equation}
while for the antisymmetric part,
\begin{equation}
\nabla \cdot (\mathbf{r} \times \vec{T}_{a}) - \mathbf{r} \times (\nabla \cdot \vec{T}_{a}) = -2\vec{T}_{a}.
\label{eq:T-asymmetric-express}
\end{equation}
In addition, we have the relatinoship of 
\begin{equation}
\bigl[\nabla \vec{u} - (\nabla \vec{u})^{\top}\bigr]_{ij}
= - \epsilon_{ijk} (\nabla \times \vec{u})_k .
\label{eq:asymveltensorandrot}
\end{equation}
Substituting Eqs.~\eqref{eq:antisymmetric-stress}--\eqref{eq:asymveltensorandrot}, Eq.~\eqref{eq:spin-evolution-raw} reduces to
\begin{equation}
\rho J \frac{D\boldsymbol{\omega}_{0}}{Dt}
= \nabla \cdot \mathbf{m}
+ 2\mu_r (\nabla \times \mathbf{u} - 2\boldsymbol{\omega}_{0})
+ \rho \mathbf{c}.
\label{eq:spin-balance-condiff}
\end{equation}
Adopting the general isotropic couple stress law of Eq.~\eqref{eq:mmeaning} with 
\[
\alpha_1 = \nu_1 + \nu_2, \quad
\alpha_2 = 0, \quad
\alpha_3 = \nu_2 + \frac{\nu_1}{3},
\]
Eq.~\eqref{eq:spin-balance-condiff} is expressed as follows:
\begin{eqnarray}
\therefore~~\rho J \frac{D\boldsymbol{\omega}_{0}}{Dt}  
= \bigl( \nu_2 + \frac{\nu_1}{3} \bigr) \nabla \nabla \cdot \boldsymbol{\omega}_{0} 
+ (\nu_1 + \nu_2) \nabla^2 \boldsymbol{\omega}_{0} 
+ 2\mu_r (\nabla \times \mathbf{u} - 2\boldsymbol{\omega}_{0}) + \rho \mathbf{c}. \label{eq:spin-condiff-expanded}
\end{eqnarray}
Here, $\nu_1$ and $\nu_2$ are the rotational moduli, which have the dimension $[{\rm M L T^{-1}}]$ in terms of mass ${\rm [M]}$, length ${\rm [L]}$, and time ${\rm [T]}$.

Finally, the balance of linear momentum given in Eq.~\eqref{eq:translational-balance} explicitly reads
\begin{equation}
\rho \frac{D\mathbf{u}}{Dt}
= \nabla \cdot \mathbf{T}_{s} + \nabla \cdot \mathbf{T}_{a} + \rho \mathbf{f}.
\label{eq:translation-final}
\end{equation}
Taking the divergence of the antisymmetric stress tensor from Eq.~\eqref{eq:antisymmetric-stress} and using Eqs.~\eqref{eq:W-definition} and \eqref{eq:asymveltensorandrot}, we obtain
\begin{align}
(\nabla \cdot \mathbf{T}_a)_i
= \partial_j (\mathbf{T}_a)_{ij} 
= \mu_r \bigl[ 2\, \partial_j \epsilon_{ijk} \omega^{0}_{k} - \epsilon_{ijk} \partial_j (\nabla \times \mathbf{u})_k \bigr], 
\end{align}
where
\begin{align}
\epsilon_{ijk} \partial_j (\nabla \times \mathbf{u})_k
= (\delta_{il}\delta_{jm} - \delta_{im}\delta_{jl}) \partial_j \partial_l u_m 
= \partial_i (\partial_j u_j) - (\nabla^2 \mathbf{u})_i. 
\end{align}
Thus we find
\begin{equation}
\nabla \cdot \mathbf{T}_{a} 
= \mu_{r} \bigl[ 2\, \nabla \times \boldsymbol{\omega}_0 
 - \nabla \nabla \cdot \mathbf{u}
 + \nabla^2 \mathbf{u} \bigr].
\label{eq:DivTa}
\end{equation}
Similarly, taking the divergence of the symmetric stress tensor from Eq.~\eqref{eq:symmetric-stress} yields
\begin{equation}
\nabla \cdot \mathbf{T}_{s} 
= -\nabla P 
+ \mu \bigl[ \nabla^2 \mathbf{u} + \nabla \nabla \cdot \mathbf{u} \bigr]
+ \lambda \nabla \nabla \cdot \mathbf{u}.
\label{eq:DivTs}
\end{equation}
Substituting Eqs.~\eqref{eq:DivTa} and \eqref{eq:DivTs} into Eq.~\eqref{eq:translation-final} gives
\begin{eqnarray}
\therefore\quad
\rho \frac{D\mathbf{u}}{Dt} 
= -\nabla P ~+~ (\mu + \mu_{r})\nabla^2 \mathbf{u} 
+ \bigl( \mu + \lambda - \mu_{r} \bigr) \nabla \nabla \cdot \mathbf{u} 
+ 2\mu_r \nabla \times \boldsymbol{\omega}_0 ~+~ \rho \mathbf{f}.
\label{eq:translation-condiff}
\end{eqnarray}
Accordingly, Eq.~\eqref{eq:translation-condiff} recovers the form of Eq.~\eqref{eq:NSwithSAMC} under the identifications
\begin{equation}
\eta = \frac{\mu}{\rho}, \quad
\eta_{r} = \frac{\mu_{r}}{\rho}, \quad
\rho \xi = \frac{2}{3}\mu + \lambda. \label{eq:etamurhoxilam}
\end{equation}
}

\HLL{The above analysis demonstrates that Condiff's Navier-Stokes equations with intrinsic spin angular momentum conservation emerge as a special case of Eringen's micropolar fluid theory, obtained by simplifying the couple stress model through the assumption $\alpha_{2} = 0$ in Eq.~\eqref{eq:mmeaning}, which neglects the effect of the transpose of the micro-rotation gradient. This assumption implies that the couple stress responds isotropically to micro-rotation gradients. In other words, it yields a model in which torsion or directional switching induced by the transposed components of the micro-rotation gradient does not occur, enforcing a linear response purely determined by the isotropic part of the micro-rotation field within the micropolar framework.}

\section{\HLL{Derivation of the Green's function with internal subgrid stress structure from micropolar fluid theory}} \label{sec:rederivmulti}
\HLL{In the time-evolution equations for the spin angular velocity derived in the previous section, we explicitly decompose the variables into large-scale and small-scale components based on turbulence theory. In this section, we derive the explicit form of the associated subgrid-scale (SGS) terms. Under the assumptions of a steady state, an isolated system, and solenoidal conditions for both the vorticity and the local angular velocity fields, we analyze the resulting mean-field equations and demonstrate that the vorticity satisfies a Poisson equation, whose Green's function intrinsically incorporates the internal structure of the subgrid stresses. We also discuss the existence of long-time stable solutions. Notably, the notion of ``multiscaleness'' invoked in Eq.~(\ref{eq:TransOmega}) in Sec.~\ref{sec:GEOFMultiFluidMech} fundamentally differs from that used in classical fluid mechanics---a point we clarify in detail. We show that the transfer of structural information across scales from small-scale features is mediated by the subgrid stresses, and that the rotational viscosity term $\nabla\times\boldsymbol{\omega}$ plays a central role in this mechanism.}

\HLL{
\subsection{Derivation of the evolution equations for vorticity and micro-rotation}
By applying the curl operator $\nabla \times$ to both sides of Eq.~\eqref{eq:translation-condiff}, and using the definition of the vorticity $\boldsymbol{\omega} = \nabla \times \mathbf{u}$, we obtain
\begin{align}
\rho \frac{D \boldsymbol{\omega}}{Dt}
= (\mu + \mu_r) \nabla^2 \boldsymbol{\omega}
+ 2\mu_r \nabla \times (\nabla \times \boldsymbol{\omega}_0) 
+ \rho \nabla \times \mathbf{f}.
\label{eq:vorticity_raw}
\end{align}
Here, we have employed the identity $\nabla^2 (\nabla \times \mathbf{u}) 
= \nabla^2 \boldsymbol{\omega}$, which can be derived in tensor notation as 
$\epsilon_{ijk} \partial_j (\partial_l \partial_l u_k)
= \partial_l \partial_l (\epsilon_{ijk} \partial_j u_k)
= \partial_l \partial_l \omega_i$. Furthermore, by using 
$\nabla \times (\nabla \times \boldsymbol{\omega}_0) 
= \nabla (\nabla \cdot \boldsymbol{\omega}_0) - \nabla^2 \boldsymbol{\omega}_0$, 
the vorticity evolution equation can be written as
\begin{align}
\rho \frac{D \boldsymbol{\omega}}{Dt}
= (\mu + \mu_r) \nabla^2 \boldsymbol{\omega} 
+ 2\mu_r \nabla \nabla \cdot \boldsymbol{\omega}_0
- 2\mu_r \nabla^2 \boldsymbol{\omega}_0  
+ \rho \nabla \times \mathbf{f}.
\label{eq:vorticity_eq}
\end{align}
On the other hand, by substituting $\boldsymbol{\omega} = \nabla \times \mathbf{u}$ into Eq.~\eqref{eq:spin-condiff-expanded}, 
the evolution equation for the local angular velocity field becomes
\begin{align}
\rho J \frac{D \boldsymbol{\omega}_0}{Dt} 
= \left( \nu_2 + \frac{\nu_1}{3} \right) \nabla \nabla \cdot \boldsymbol{\omega}_0
+ (\nu_1 + \nu_2) \nabla^2 \boldsymbol{\omega}_0 
+ 2\mu_r (\boldsymbol{\omega} - 2\boldsymbol{\omega}_0)
+ \rho \mathbf{c}.
\label{eq:micro_rotation_eq}
\end{align}
In these equations, $\rho$ denotes the fluid density, $\mathbf{u}$ the velocity field, $P$ the pressure, and $\mathbf{f}$ the external body force. The vector $\boldsymbol{\omega}_0$ represents the local angular velocity (micro-rotation), and $\mathbf{c}$ denotes the local couple stress. The shear viscosity is denoted by $\mu$, and the rotational viscosity by $\mu_r$. The local angular moment of inertia is given by $\rho J$, and the rotational moduli are denoted by $\nu_1$ and $\nu_2$. The operator $D/Dt = \partial/\partial t + (\mathbf{u} \cdot \nabla)$ denotes the material derivative following the fluid element.

\subsection{Scale separation and filtering}
Next, based on the concept of large eddy simulation (LES), we explicitly separate the variables into large-scale and small-scale components:
\begin{align}
\boldsymbol{\omega} = \overline{\boldsymbol{\omega}} + \boldsymbol{\omega}', \quad
\boldsymbol{\omega}_0 = \overline{\boldsymbol{\omega}_0} + \boldsymbol{\omega}_0', \quad
\mathbf{u} = \overline{\mathbf{u}} + \mathbf{u}',
\label{eq:filteringval}
\end{align}
where $\overline{(\cdot)}$ denotes a filtering operation, typically represented by a spatial average using a Gaussian kernel $G$ of width $\Delta$. Under conditions where solid body rotation is always satisfied, Eqs.~\eqref{eq:vorticity_eq} and \eqref{eq:micro_rotation_eq} should be exactly equivalent. By substituting $\boldsymbol{\omega} = 2\boldsymbol{\omega}_0$ into Eq.~\eqref{eq:filteringval}, we have
\begin{equation}
\overline{\boldsymbol{\omega}} + \boldsymbol{\omega}' = 2\overline{\boldsymbol{\omega}_0}+2\boldsymbol{\omega}_0'.
\end{equation}
Assuming this equality holds for each corresponding physical quantity, it is natural to impose
\begin{align}
\overline{\boldsymbol{\omega}}=2\overline{\boldsymbol{\omega}_0},\quad
\boldsymbol{\omega}'=2\boldsymbol{\omega}_0'. 
\label{eq:solidcondeach}
\end{align}
By substituting Eq.~\eqref{eq:filteringval} into Eqs.~\eqref{eq:vorticity_eq} and \eqref{eq:micro_rotation_eq}, and applying the relation $\boldsymbol{\omega}'=2\boldsymbol{\omega}_0'$, the mean-field equations become
\begin{align}
\rho \left( 
\frac{\partial \overline{\boldsymbol{\omega}}}{\partial t}
+ (\overline{\mathbf{u}} \cdot \nabla) \overline{\boldsymbol{\omega}}
\right)
= (\mu + \mu_r) \nabla^2 \overline{\boldsymbol{\omega}} 
+ 2\mu_r \nabla \nabla \cdot \overline{\boldsymbol{\omega}_0}
- 2\mu_r \nabla^2 \overline{\boldsymbol{\omega}_0} 
+ \rho \nabla \times \overline{\mathbf{f}}
- \rho \boldsymbol{\tau}_\omega,
\label{eq:les_vorticity}
\end{align}
\begin{align}
\rho J 
\left( 
\frac{\partial \overline{\boldsymbol{\omega}_0}}{\partial t}
+ (\overline{\mathbf{u}} \cdot \nabla) \overline{\boldsymbol{\omega}_0}
\right)
= \left( \nu_2 + \frac{\nu_1}{3} \right) \nabla \nabla \cdot \overline{\boldsymbol{\omega}_0} 
+ (\nu_1 + \nu_2) \nabla^2 \overline{\boldsymbol{\omega}_0} 
+ 2\mu_r (\overline{\boldsymbol{\omega}} - 2\overline{\boldsymbol{\omega}_0}) 
+ \rho \overline{\mathbf{c}}
- \rho J \boldsymbol{\tau}_{\omega_0}.
\label{eq:les_omega0}
\end{align}
Here, the subgrid-scale (SGS) terms emerge from the averaging of the material derivatives. Specifically, the difference in the advective term of the vorticity is defined as
\begin{align}
\boldsymbol{\tau}_\omega 
= \overline{ (\mathbf{u} \cdot \nabla) \boldsymbol{\omega}} 
- (\overline{\mathbf{u}} \cdot \nabla) \overline{\boldsymbol{\omega}},
\label{eq:sgs_omega}
\end{align}
and similarly for the angular velocity field,
\begin{align}
\boldsymbol{\tau}_{\omega_0}
= \overline{ (\mathbf{u} \cdot \nabla) \boldsymbol{\omega}_0} 
- (\overline{\mathbf{u}} \cdot \nabla) \overline{\boldsymbol{\omega}_0}.
\label{eq:sgs_omega0}
\end{align}
In deriving Eqs.~\eqref{eq:les_vorticity} and \eqref{eq:les_omega0}, we have assumed the statistical averaging conditions on the small scales:
\begin{align}
\frac{\overline{\partial \boldsymbol{\omega}'}}{\partial t} = 0,\quad
\frac{\overline{\partial \boldsymbol{\omega}_0'}}{\partial t} = 0, \quad
\overline{\nabla^{2}\boldsymbol{\omega'}} = 0,\quad
\overline{\nabla^{2}\boldsymbol{\omega}_0'} = 0.
\end{align}
Moreover, by substituting $\boldsymbol{\omega}'=2\boldsymbol{\omega}_0'$ into Eqs.~\eqref{eq:sgs_omega} and \eqref{eq:sgs_omega0}, we obtain
\begin{equation}
\boldsymbol{\tau}_{\omega} = 2\boldsymbol{\tau}_{\omega_0}.
\label{eq:taueqtwotau0}
\end{equation}

For simplicity, we now assume a steady state (or a long-time average) and neglect external forces and couples. Under the solenoidal conditions $\nabla\cdot \boldsymbol{\omega}=\nabla\cdot \boldsymbol{\omega}_{0} =0$, the solid rotation condition $\overline{\boldsymbol{\omega}}=2\overline{\boldsymbol{\omega}_0}$, and Eq.~\eqref{eq:taueqtwotau0}, Eqs.~\eqref{eq:les_vorticity} and \eqref{eq:les_omega0} reduce to
\begin{eqnarray}
    (\overline{\mathbf{u}} \cdot \nabla) \overline{\boldsymbol{\omega}} &=& \frac{\mu}{\rho} \nabla^2 \overline{\boldsymbol{\omega}} - \boldsymbol{\tau}_\omega, \label{eq:steady_vorticity} \\[1ex]  
    (\overline{\mathbf{u}} \cdot \nabla) \overline{\boldsymbol{\omega}} &=& \frac{(\nu_1 + \nu_2)}{\rho J } \nabla^2 \overline{\boldsymbol{\omega}} - \boldsymbol{\tau}_\omega. \label{eq:steady_omega0}
\end{eqnarray}
By comparing the corresponding terms in Eqs.~\eqref{eq:steady_vorticity} and \eqref{eq:steady_omega0}, we find
\begin{equation}
    \therefore \mu = \frac{\nu_1 + \nu_2}{J}.
\label{eq:v1v2Jbreak}
\end{equation}
If we take the dimensions of mass $[\mathrm{M}]$, length $[\mathrm{L}]$, and time $[\mathrm{T}]$, then $\nu_1$ and $\nu_2$ are moduli with dimensions $[\mathrm{MLT^{-1}}]$, and the moment of inertia per unit volume $J$ has dimensions $[\mathrm{M}^2]$. Thus, we can verify that $\mu$ has dimensions $[\mathrm{M L^{-1} T^{-1}}]$, which is consistent with the usual viscosity.

Equation~\eqref{eq:v1v2Jbreak} represents a crucial relation that connects the coefficients associated with micro-rotation to the macroscopic transport coefficients. Both Eq.~\eqref{eq:steady_vorticity} and Eq.~\eqref{eq:steady_omega0} are mean-field equations for the vorticity, derived on the basis of micropolar fluid theory. However, Eq.~\eqref{eq:steady_vorticity} is obtained from the evolution equation for the velocity field $\mathbf{u}$, while Eq.~\eqref{eq:steady_omega0} is derived from the evolution equation for the micro-rotation (local angular velocity) of material points (or fluid particles), and thus involves microscopic parameters such as the moment of inertia $J$ and the rotational moduli $\nu_1$ and $\nu_2$.

Since material points or fluid particles are inherently hypothetical constructs, it is not straightforward to estimate the parameters associated with micro-rotation. From an ontological perspective, these parameters must be determined from the microscopic material properties of the molecules or atoms that constitute the continuum. However, due to fundamental differences in the governing mechanical laws, this is generally a challenging task. It is precisely because we derived the vorticity evolution equations from two distinct approaches within the framework of micropolar fluid theory that we were able to obtain the clear and direct relation given by Eq.~\eqref{eq:v1v2Jbreak}, which links the physical viscosity $\mu$ to the parameters characterizing micro-rotation.

\subsection{Steady solutions}
By introducing the kinematic viscosity $\eta = \mu/\rho$ and substituting 
$\overline{ (\mathbf{u} \cdot \nabla) \boldsymbol{\omega} } = -\boldsymbol{S}(\mathbf{r})$ 
along with 
$\boldsymbol{\tau}_\omega 
= \overline{ (\mathbf{u} \cdot \nabla) \boldsymbol{\omega}} 
- (\overline{\mathbf{u}} \cdot \nabla) \overline{\boldsymbol{\omega}}$
into Eq.~\eqref{eq:steady_vorticity}, we obtain
\begin{equation}
\nabla^2 \overline{\boldsymbol{\omega}}(\mathbf{r})
= -\frac{1}{\eta} \boldsymbol{S}(\mathbf{r}).
\label{eq:poissonnobound}
\end{equation}
This equation represents a typical Poisson equation for each component of $\overline{\boldsymbol{\omega}}$, 
and an integral solution can be obtained by employing the Green's function for the unbounded domain $\mathbb{R}^3$.
The fundamental solution (Green's function) $G(\mathbf{r}, \mathbf{r}')$ for the Poisson equation in $\mathbb{R}^3$ is given by
\begin{equation}
G(\mathbf{r}, \mathbf{r}') = -\frac{1}{4\pi |\mathbf{r} - \mathbf{r}'|},
\end{equation}
and thus, by utilizing this Green's function, the solution is expressed as
\begin{align}
\therefore \overline{\boldsymbol{\omega}}(\mathbf{r})
= \frac{1}{4\pi \eta}
\int_{\mathbb{R}^3} 
\frac{\boldsymbol{S}(\mathbf{r}')}{|\mathbf{r} - \mathbf{r}'|} 
\, d\mathbf{r}'.
\label{eq:omegaconvol_withs_nb}
\end{align}
In this way, the large-scale vorticity $\overline{\boldsymbol{\omega}}$ is determined by the source term 
$\boldsymbol{S}(\mathbf{r}) = -\overline{ (\mathbf{u} \cdot \nabla) \boldsymbol{\omega}}$.

\subsection{Convergence from time-dependent solutions}
The source term $\boldsymbol{S}(\mathbf{r}) = -\overline{ (\mathbf{u} \cdot \nabla) \boldsymbol{\omega}}$ originates from the advection term, and therefore has dimensions of $\mathrm{[T^{-2}]}$. On the other hand, $\overline{\boldsymbol{\omega}}$ has dimensions of $\mathrm{[T^{-1}]}$, implying that the kernel $G$ must carry temporal dimensions. Although the convolution integral involving $G$ is performed over spatial coordinates, the underlying physical process necessarily evolves over time, which is consistent with this interpretation. 

Of course, to rigorously incorporate time dependence, one must solve the Poisson equation with explicit temporal terms included. Thus, the Green's function solution presented earlier can be understood as an approximate solution that, while intrinsically dependent on temporal dimensions, effectively neglects the explicit time dependence under steady-state conditions. Up to this point, we have assumed a steady state and ignored temporal variations; however, when time dependence is taken into account, the vorticity evolution equation replaces the Poisson equation and is expressed as
\begin{equation}
\frac{\partial}{\partial t} \overline{\boldsymbol{\omega}}(\mathbf{r}, t)
= \eta \nabla^2 \overline{\boldsymbol{\omega}}(\mathbf{r}, t)
+ \boldsymbol{S}(\mathbf{r}, t).
\label{eq:time_evolution_withS}
\end{equation}
This represents a time-dependent vorticity transport equation that includes both diffusion and a source term. 
The solution to Eq.~\eqref{eq:time_evolution_withS} can be written in terms of the heat kernel function $G(\mathbf{r}-\mathbf{r'}, t)$ as
\begin{align}
\overline{\boldsymbol{\omega}}(\mathbf{r}, t)
= \int_{\mathbb{R}^3} G(\mathbf{r}-\mathbf{r'}, t) \overline{\boldsymbol{\omega}}(\mathbf{r'},0)\, d\mathbf{r'} + \int_0^t \int_{\mathbb{R}^3} G(\mathbf{r}-\mathbf{r'}, t-\tau)\, \boldsymbol{S}(\mathbf{r'}, \tau)\, d\mathbf{r'}\, d\tau,
\label{eq:heat_solution}
\end{align}
where $G(\mathbf{r}-\mathbf{r'}, t-\tau)$ is given by
\begin{equation}
G(\mathbf{r}-\mathbf{r'}, t)
= \frac{1}{(4\pi \eta t)^{3/2}}
\exp\!\left(-\frac{|\mathbf{r} - \mathbf{r'}|^2}{4\eta t}\right).
\label{eq:heat_kernel}
\end{equation}

Next, we consider the steady limit by taking $t\to\infty$. In this limit, the initial term in Eq.~\eqref{eq:heat_solution} vanishes, and assuming a time-independent $\boldsymbol{S}$, the steady solution becomes
\begin{equation}
\overline{\boldsymbol{\omega}}_{\infty}(\mathbf{r})
= \int_0^\infty \int_{\mathbb{R}^3} G(\mathbf{r}-\mathbf{r'}, \tau)\, \boldsymbol{S}(\mathbf{r'})\, d\mathbf{r'}\, d\tau.
\label{eq:steady_solution}
\end{equation}
Defining
\begin{equation}
\Phi(\mathbf{r}-\mathbf{r'}) 
:= \int_0^\infty G(\mathbf{r}-\mathbf{r'}, \tau)\, d\tau,
\label{eq:phi_def}
\end{equation}
Eq.~\eqref{eq:steady_solution} can be rewritten as
\begin{equation}
\overline{\boldsymbol{\omega}}_{\infty}(\mathbf{r})
= \int_{\mathbb{R}^3} \Phi(\mathbf{r}-\mathbf{r'})\, \boldsymbol{S}(\mathbf{r'})\, d\mathbf{r'}.
\label{eq:steady_solution_phi}
\end{equation}
From Eqs.~\eqref{eq:heat_kernel} and \eqref{eq:phi_def}, we obtain
\begin{equation}
\Phi(\mathbf{r}-\mathbf{r'})
= \int_0^\infty \frac{1}{(4\pi \eta \tau)^{3/2}}
\exp\!\left(-\frac{d^2}{4\eta \tau}\right) \, d\tau,
\label{eq:phi_integral_original}
\end{equation}
where $d = |\mathbf{r} - \mathbf{r'}|$. By making the change of variables $s = \frac{d^2}{4\eta \tau}$, we find
\begin{align}
\Phi(\mathbf{r}-\mathbf{r'})
= \frac{1}{4\eta (\pi d^2)^{3/2}} d^2
\int_0^\infty s^{-1/2} e^{-s}\, ds.
\label{eq:phi_gamma_integral}
\end{align}
Using the integral formula for the Gamma function, $\int_0^\infty s^{-1/2} e^{-s}\, ds = \Gamma\!\left(\frac12\right)= \sqrt{\pi}$, we finally arrive at
\begin{equation}
\therefore \overline{\boldsymbol{\omega}}_{\infty}(\mathbf{r})
= \frac{1}{4\pi \eta}
\int_{\mathbb{R}^3} 
\frac{\boldsymbol{S}(\mathbf{r'})}{|\mathbf{r} - \mathbf{r'}|} 
\, d\mathbf{r'},
\label{eq:poisson_solution}
\end{equation}
thus recovering the Poisson-type integral solution.
In this way, by taking the long-time average, we recover a Poisson-type Green's function solution.

Let us now discuss the correspondence between the meaning of Eq.~(\ref{eq:TransOmega}) and Eq.~\eqref{eq:poisson_solution}. 
The coefficient $1/(4\pi\eta)$ in the Green's function solution Eq.~\eqref{eq:poisson_solution} is not uniquely determined. Similarly, in Eq.~(\ref{eq:TransOmega}), the coefficient $C \in \mathbb{R}$ remains unrestricted. 
To facilitate comparison, we can rewrite Eq.~(\ref{eq:TransOmega}) as
\begin{equation}
\therefore \overline{\boldsymbol{\omega}}(\mathbf{r})
= \frac{1}{4\pi\eta}
\int_{\mathbb{R}^3} 
\frac{\zeta \boldsymbol{\omega}(\mathbf{r'})}{|\mathbf{r} - \mathbf{r'}|} 
\, d\mathbf{r'},
\label{eq:redefinedconv}
\end{equation}
where $(\zeta/2\pi\eta)^{-1}$ corresponds to the coefficient $C$ in Eq.~(\ref{eq:TransOmega}). Equation~\eqref{eq:redefinedconv} corresponds to the steady-state approximation of the time-dependent vorticity transport equation with diffusion and source terms, as given by Eq.~\eqref{eq:poisson_solution}. A formal comparison between the two reveals that $\zeta \boldsymbol{\omega}(\mathbf{r'}) = \boldsymbol{S}(\mathbf{r'})$, with the right-hand side encapsulating the information from the small-scale vorticity. It is important to note here that both sides of Eq.~\eqref{eq:redefinedconv} represent the same physical quantity (vorticity). Thus, unlike Eq.~\eqref{eq:poisson_solution}, the convolution integral in Eq.~\eqref{eq:redefinedconv} carries no explicit temporal dimension, rendering it dimensionless. 

Equation~\eqref{eq:redefinedconv} can be interpreted as representing a mapping from $\boldsymbol{\omega}$ to $\overline{\boldsymbol{\omega}}$. This perspective highlights the difference between the notion of scale underlying Eq.~\eqref{eq:redefinedconv} and the concept of multiscaleness commonly employed in classical turbulence studies such as LES. If we denote by $g$ a mapping from large to small scales, then in the present case we have $g:\overline{\boldsymbol{\omega}}\rightarrow \boldsymbol{\omega}$, whereas in classical LES it would be expressed as $g:\overline{\boldsymbol{\omega}}\rightarrow \boldsymbol{\omega}'$. Namely, the former represents a transformation to a quantity that inherently contains small-scale information, while the latter transforms directly to the small-scale quantity itself. In fact, the discussion on multiscaleness in Sec.~\ref{sec:GEOFMultiFluidMech} can also be interpreted entirely in the context of such mapping spaces (whether viewed as a physical integral or as a state-to-state mapping). Even under this interpretation, the nested structure seen in Eqs.~(\ref{eq:RecurrRel}) and (\ref{eq:GeneralRecurrRel}) remains valid as a mathematical property, and thus the same explanation applies. Specifically, the vorticity on the left-hand side of Eq.~(\ref{eq:RecurrRel}) traces through the second term on the right-hand side and, via the structure of $\nabla\times\boldsymbol{\omega}$, ultimately connects to finer-scale (high-frequency) components of the vorticity. The physical picture of such high-frequency vortices has already been discussed: namely, since no vortices exist below the quantum vortex scale, the second, recursively defined term on the right-hand side of Eq.~(\ref{eq:RecurrRel}) vanishes, effectively terminating the nested structure. Although vortices below the Kolmogorov scale are typically dissipated by molecular viscosity and thus do not persist, as discussed in Section~2, such residual structures---like quantum vortices---may be treated as the base scale in this context. Accordingly, regardless of the physical interpretation of the convolution-based transformation, the structure of scale-to-scale information transfer via $\nabla\times\boldsymbol{\omega}$ remains unchanged and can be meaningfully asserted.

While it is common to impose the condition that the integral of the vorticity over the entire space vanishes---for instance, to ensure the uniqueness of the velocity field---this is not a fundamental physical constraint. In many cases, the spatial integral of vorticity, particularly along a given direction, carries physical significance. A notable example is rotating cryogenic liquid helium-4, where the total circulation along the rotation axis is quantized and corresponds to a nonzero integral of vorticity~\cite{Yarmchuk1982, PhysRevLett.86.4443}. This demonstrates that net vorticity in a specific direction can represent a physically meaningful quantity, rather than merely a mathematical artifact. Nevertheless, for Eq.~\eqref{eq:redefinedconv} to provide a physically meaningful description, it must correspond precisely to Eq.~\eqref{eq:poisson_solution}. There also remains the issue of the mathematical stability of the solution, which we address next. For the two equations to be equivalent, it is necessary that $\zeta \boldsymbol{\omega} = \boldsymbol{S}$ holds. The right-hand side, $\boldsymbol{S}$, is given by $-\overline{ (\mathbf{u} \cdot \nabla) \boldsymbol{\omega}}$ and represents a nonlinear term aggregating unknowns. Determining this term is generally referred to as the closure problem; it is difficult to derive $\boldsymbol{S}$ purely from first principles, and modeling is typically required. Accordingly, we shall first examine the stability of the solution under the mathematical assumption that $\zeta \boldsymbol{\omega} = \boldsymbol{S}$ holds.

\subsection{Stability analysis of the solution}
By substituting the relation $\zeta \boldsymbol{\omega}(\mathbf{r'}) = \boldsymbol{S}(\mathbf{r'})$ into Eq.~\eqref{eq:time_evolution_withS}, we obtain
\begin{equation}
\partial_t \overline{\boldsymbol{\omega}}(\mathbf{r}, t)
- \eta \nabla^2 \overline{\boldsymbol{\omega}}(\mathbf{r}, t)
= \zeta \boldsymbol{\omega}(\mathbf{r}, t).
\label{eq:main_equation}
\end{equation}
In the following, we examine in detail the stability and existence conditions of solutions to Eq.~\eqref{eq:main_equation} based on the parameter $\zeta$. Here, $\eta > 0$ denotes the kinematic viscosity, $\zeta$ is a parameter corresponding to the growth or reaction rate of the system, and $\boldsymbol{\omega}$ represents the vorticity field. To analyze the long-time behavior, we approximate the right-hand side of Eq.~\eqref{eq:main_equation} by $\boldsymbol{\omega} \approx \overline{\boldsymbol{\omega}}$. Let us consider the case where the initial value $\overline{\boldsymbol{\omega}}(\vec{r}, 0)$ is known constant, which is denoted by $\hat{\boldsymbol{\omega}}(\vec{k}, 0)$ in wevenumber space. In this case, Eq.~\eqref{eq:main_equation} is a linear constant-coefficient diffusion-reaction type equation. By applying a Fourier mode expansion:
\[
\overline{\boldsymbol{\omega}}(\mathbf{r}, t)
= \int_{\mathbb{R}^3} \hat{\omega}(\mathbf{k}, t) e^{i\mathbf{k}\cdot\mathbf{r}} \, d\mathbf{k},
\]
it reduces for each mode to
\begin{align}
\partial_t \hat{\omega}(\mathbf{k}, t)
= -\eta |\mathbf{k}|^2 \hat{\omega}(\mathbf{k}, t)
+ \zeta \hat{\omega}(\mathbf{k}, t)  
= \left(\zeta - \eta |\mathbf{k}|^2\right) \hat{\omega}(\mathbf{k}, t).
\label{eq:fourier_ode}
\end{align}
The solution for each mode with initial condition $\hat{\omega}(\mathbf{k}, 0)$ is given by
\begin{equation}
\hat{\omega}(\mathbf{k}, t)
= \hat{\omega}(\mathbf{k}, 0) \, \exp\!\left[\bigl(\zeta - \eta |\mathbf{k}|^2\bigr) t\right].
\label{eq:fourier_solution}
\end{equation}
From this behavior, we can classify the solutions according to the sign of $\zeta$ as follows.

\bigskip

(i) When $\zeta < 0$, we have
\[
\zeta - \eta |\mathbf{k}|^2 < 0
\]
for all modes. Thus, all Fourier components decay exponentially over time. Consequently, for any initial condition, the solution $\overline{\boldsymbol{\omega}}(\mathbf{r}, t)$ tends to zero as $t \to \infty$, and its $L^2$ norm remains bounded. In this case, Eq.~\eqref{eq:main_equation} is stable, and the steady field vanishes in the infinite-time limit.

\bigskip

(ii) When $\zeta = 0$, Eq.~\eqref{eq:fourier_solution} simplifies to
\[
\hat{\omega}(\mathbf{k}, t)
= \hat{\omega}(\mathbf{k}, 0) \, \exp\!\left[-\eta |\mathbf{k}|^2 t\right].
\]
Here, the zero mode with $|\mathbf{k}|=0$ (the mean component) remains constant in time, while all higher modes decay due to diffusion. As a result, the global solution retains its initial mean component, with all other structures smoothed out by diffusion. In this case, the solution remains bounded for all time and is stable.

\bigskip

(iii) When $\zeta > 0$, low-wavenumber modes with $|\mathbf{k}|^2 < \frac{\zeta}{\eta}$ satisfy
\[
\zeta - \eta |\mathbf{k}|^2 > 0,
\]
which leads to exponential growth over time. Therefore, for any initial disturbance, the solution grows without bound as time progresses, and its $L^2$ norm diverges. In this scenario, Eq.~\eqref{eq:main_equation} is unstable, and globally bounded solutions do not exist.

\bigskip

In summary, the stability and existence of solutions to Eq.~\eqref{eq:main_equation} are determined by the sign of the parameter $\zeta$. In particular, when $\zeta \le 0$, globally bounded solutions exist for all time under arbitrary initial conditions, with all modes decaying exponentially if $\zeta < 0$. Conversely, when $\zeta > 0$, growth modes appear that surpass the stabilizing effect of diffusion, leading to arbitrarily large solutions over long times and a loss of stability. On the other hand, since this is essentially a typical Cauchy problem, a solution always exists instantaneously regardless of the value of $\zeta$. Nevertheless, from the definition in Eq.~(\ref{eq:TransOmega}), it is unlikely that $\zeta=0$, and in practice, we expect that $\zeta \in \mathbb{R}$ with $\zeta \ne 0$, allowing variations in its value provided they do not exert a dominant influence on the integral behavior.

Intuitively, the parameter $\zeta$ can be regarded as controlling the tendency of the vorticity field to grow or decay, and it may be loosely interpreted as analogous to the sign (or direction) of vorticity. The analysis in this section indicates that, although the long-time behavior and impact on the global system differ depending on the sign of $\zeta$ (as seen in cases (i) and (iii)), the Cauchy nature of the problem guarantees the local existence of solutions regardless of the sign, thereby allowing initial conditions with arbitrary sign to contribute to the global integral representation of the solution. It should be noted, however, that the above analysis is based on a linear approximation. In particular, the exponential growth observed when $\zeta > 0$ may be suppressed in practice by nonlinear effects, boundary conditions, or energy conservation constraints. Therefore, even though linear stability analysis in an idealized infinite domain may predict instability, actual systems may exhibit saturation or transitions to new steady states. Such nonlinear effects should be examined through more detailed modeling and numerical analysis.

\subsection{Modeling and dimensional justification of the vorticity source term}
Next, we examine the validity of modeling the source term in the vorticity transport equation,
\begin{equation}
\boldsymbol{S}(\mathbf{r})
= -\overline{ (\mathbf{u} \cdot \nabla) \boldsymbol{\omega} },
\label{eq:source_term}
\end{equation}
by the closure relation
\begin{equation}
\boldsymbol{S}(\mathbf{r}) = \zeta\, \boldsymbol{\omega}(\mathbf{r}),
\label{eq:closure_model}
\end{equation}
where $\zeta$ is a scaling parameter with the dimension of inverse time ($[\zeta] = T^{-1}$), representing the characteristic time scale associated with the evolution of vorticity. Equation~\eqref{eq:source_term} arises from averaging the nonlinear advection term $(\mathbf{u}\cdot\nabla)\boldsymbol{\omega}$ in the fluid equations, and is a typical feature encountered in closure problems such as those in LES and RANS. However, it cannot be strictly determined by the mean field $\overline{\boldsymbol{\omega}}$ or the filtered large-scale variables alone, and it is therefore impossible to close it purely from axiomatic principles. Thus, some form of physical or statistical approximation must be introduced. Equation~\eqref{eq:closure_model} represents the simplest linear approximation model, and its validity can be argued on the following grounds.

First, from a dimensional analysis standpoint, the dimensions of Eq.~\eqref{eq:source_term} are given by
\begin{equation}
[\boldsymbol{S}] = [\mathbf{u}] [\nabla] [\boldsymbol{\omega}]
= \frac{L}{T} \cdot \frac{1}{L} \cdot \frac{1}{T} = \frac{1}{T^2},
\label{eq:dimensional_S}
\end{equation}
while the right-hand side of Eq.~\eqref{eq:closure_model} has dimensions
\begin{equation}
[\zeta \boldsymbol{\omega}]
= \frac{1}{T} \cdot \frac{1}{T} = \frac{1}{T^2},
\label{eq:dimensional_closure}
\end{equation}
showing perfect dimensional consistency.

Next, under long-time averages or statistically steady states in turbulent flows, it has been frequently observed that the vorticity field, which is nonlinearly advected over short time scales, statistically relaxes and redistributes in proportion to the local mean vorticity. This behavior can be interpreted as a form of linear response to the coarse-grained mean vorticity field. Recent analyses have reported that vortex fluctuations, when examined in a scale-separated framework, exhibit a statistical alignment with the filtered mean vorticity, suggesting a near-linear proportionality between the conditional vorticity $\omega'$ and its local mean $\overline{\omega}$ \cite{JFM_Tobias,PhysRevE_Wilczek,PhysRevFluids_Buaria}. In particular, the study in \cite{JFM_Tobias} demonstrates, within a linear response theory for vortex meandering, that long-time fluctuations of vorticity around $\overline{\omega}$ follow a Langevin-type dynamics whose ensemble-averaged behavior is proportional to the mean field. This is consistent with earlier works highlighting the statistical organization of nonlinearly advected vorticity under coarse-graining in turbulent flows \cite{PhysRevE_Wilczek,PhysRevFluids_Buaria}. 

Thus, Eq.~\eqref{eq:closure_model} naturally emerges as the simplest proportional model for the mean field, reflecting the statistical organization of vorticity observed in turbulent flows. Building on this interpretation, it is instructive to recall that many subgrid-scale (SGS) closures in LES and RANS adopt a similar principle of linear response, often implemented through an effective eddy viscosity. Specifically, the nonlinear advection term is frequently modeled in the form
\[
- \overline{ (\mathbf{u} \cdot \nabla) \boldsymbol{\omega} }
\approx -\nu_t \nabla^2 \overline{\boldsymbol{\omega}},
\]
which acts as a diffusive closure. By locally simplifying this representation even further, one arrives at
\[
- \overline{ (\mathbf{u} \cdot \nabla) \boldsymbol{\omega} }
\approx \zeta\, \overline{\boldsymbol{\omega}},
\]
so that Eq.~\eqref{eq:closure_model} can be viewed as an extreme limiting approximation of this class of models. Therefore, although Eq.~\eqref{eq:closure_model} cannot be derived purely from first principles, it can be considered a reasonable modeling choice in light of dimensional consistency, physical interpretation, and established mean-field and LES modeling frameworks. This type of linear proportional closure is conceptually consistent with the Boussinesq hypothesis commonly employed in Reynolds-averaged turbulence models, where turbulent fluxes or stresses are modeled as linearly related to mean field gradients. While Eq.~\eqref{eq:closure_model} is introduced heuristically, it shares this foundational idea of representing unresolved dynamics through a linear response to resolved quantities.

\subsection{Scale transfer mediated by rotational viscosity}
Finally, we derive the form that characterizes the scale-to-scale information transfer inherent in the nonlinear advection term,
\begin{equation}
\boldsymbol{S}(\mathbf{r})
= -\overline{ (\mathbf{u} \cdot \nabla) \boldsymbol{\omega} }.
\label{eq:original_S}
\end{equation}
For simplicity, we assume the incompressible condition $\nabla\cdot\mathbf{u}=0$ in the following discussion. 

By applying the vector identity
\[
(\mathbf{u}\cdot\nabla)\boldsymbol{\omega}
= \nabla(\mathbf{u}\cdot\boldsymbol{\omega})
- \mathbf{u}\times(\nabla\times\boldsymbol{\omega})
- \boldsymbol{\omega}\times(\nabla\times\mathbf{u})
\]
together with the incompressibility condition $\nabla\cdot\mathbf{u}=0$, the definition of vorticity $\nabla\times\mathbf{u}=\boldsymbol{\omega}$, and the identity $\boldsymbol{\omega}\times\boldsymbol{\omega}=0$, we obtain
\begin{equation}
(\mathbf{u}\cdot\nabla)\boldsymbol{\omega}
= \nabla(\mathbf{u}\cdot\boldsymbol{\omega})
- \mathbf{u}\times(\nabla\times\boldsymbol{\omega}).
\label{eq:identity_simplified}
\end{equation}
Thus, the nonlinear advection term becomes
\begin{equation}
\boldsymbol{S}(\mathbf{r})
= -\overline{\nabla(\mathbf{u}\cdot\boldsymbol{\omega})}
+ \overline{\mathbf{u}\times(\nabla\times\boldsymbol{\omega})}.
\label{eq:S_curlform}
\end{equation}

Now, if the velocity field decays sufficiently at infinity (or periodic boundary conditions are imposed), the divergence theorem yields
\[
\int_{\mathbb{R}^3} \overline{\nabla(\mathbf{u}\cdot\boldsymbol{\omega})}\, d\mathbf{r}
= \oint_{\partial \Omega} \overline{(\mathbf{u}\cdot\boldsymbol{\omega})} \, \mathbf{n}\, dS
= 0.
\]
Hence, this term does not contribute globally, and the primary contribution to the local vorticity transport is given by
\begin{equation}
\boldsymbol{S}(\mathbf{r})
= \overline{\mathbf{u}\times(\nabla\times\boldsymbol{\omega})}.
\label{eq:S_rotation}
\end{equation}
Furthermore, by expressing $\mathbf{u}$ using the Biot-Savart representation,
\begin{equation}
\mathbf{u}(\mathbf{r})
= \frac{1}{4\pi} \int_{\mathbb{R}^3}
\frac{(\mathbf{r}-\mathbf{r}')\times \boldsymbol{\omega}(\mathbf{r}')}{|\mathbf{r}-\mathbf{r}'|^3}\, d\mathbf{r}',
\label{eq:biot_savart}
\end{equation}
Eq.~\eqref{eq:S_rotation} can be written entirely as a double integral over the vorticity field $\boldsymbol{\omega}$,
\begin{equation}
\therefore \boldsymbol{S}(\mathbf{r})
= \frac{1}{4\pi} \overline{
\int_{\mathbb{R}^3}
\left[
\frac{(\mathbf{r}-\mathbf{r}')\times \boldsymbol{\omega}(\mathbf{r}')}{|\mathbf{r}-\mathbf{r}'|^3}
\right]
\times (\nabla\times\boldsymbol{\omega}(\mathbf{r}))
\, d\mathbf{r}'
}.
\label{eq:S_doubleintegral}
\end{equation}
This expression illustrates how the rotational viscosity structure $\nabla\times\boldsymbol{\omega}$ at the point $\mathbf{r}$ is nonlocally influenced by the vorticity field $\boldsymbol{\omega}(\mathbf{r}')$ throughout the entire space, and conversely induces local transport and dissipation. In other words, it explicitly captures the mechanism of direct scale-to-scale information transfer between small-scale structures and large-scale fields.}

\HLL{To elaborate further, the term characterizing rotational viscosity, $\nabla\times\boldsymbol{\omega}$, depends on higher-order spatial structures compared to the conventional vorticity transport, and thus exhibits particularly pronounced nonlocality. In fact, in Fourier space, $\nabla\times\boldsymbol{\omega}$ is represented as $i\mathbf{k}\times\hat{\omega}(\mathbf{k})$, which increases in magnitude proportional to the wavenumber $|\mathbf{k}|$~($=k$), indicating that it preferentially captures high-wavenumber (small-scale) components. More precisely, since vorticity is given by $\boldsymbol{\omega} = \nabla \times \mathbf{u}$, it scales linearly with the wavenumber $k$ in Fourier space, so that $|\boldsymbol{\omega}|^2$ carries a weight of approximately $\sim k^2$ in the energy spectrum, emphasizing higher wavenumber (i.e., smaller scale) features. On the other hand, $\nabla \times \boldsymbol{\omega}$ involves yet another spatial derivative, leading to a scaling of $\sim k^2$ relative to $\boldsymbol{\omega}$, and consequently $|\nabla \times \boldsymbol{\omega}|^2$ carries a weight of $\sim k^4$. 

As a result, while $|\boldsymbol{\omega}|^2$ is somewhat sensitive to small-scale structures, $|\nabla \times \boldsymbol{\omega}|^2$ is even more sensitive to higher wavenumbers, i.e., finer spatial structures, and thus can be regarded as depending on higher-order spatial features. Furthermore, when this term couples into the transport term through the Biot--Savart integral representation of the velocity field, the local rotational viscosity structure $\nabla\times\boldsymbol{\omega}$ becomes nonlocally linked with the vorticity field $\boldsymbol{\omega}(\mathbf{r}')$ throughout the entire space, as explicitly formulated in Eq.~\eqref{eq:S_doubleintegral}. This equation illustrates the mechanism by which the rotational viscosity structure at point $\mathbf{r}$ is influenced nonlocally by the global vorticity field, resulting in local transport and rearrangement. Consequently, this facilitates information transfer across a broader range of scales than would occur via mere vortex-to-vortex transport, establishing pathways by which small-scale vortex structures can be directly connected to distant large-scale structures. Therefore, the inclusion of the rotational viscosity term in the transport equation can be interpreted as a primary physical mechanism by which small-scale structures are dynamically transported and reorganized over wider spatial scales.

In Section~\ref{sec:simulation}, we will consider a fundamental benchmark problem of freely decaying two-dimensional turbulence starting from multiple point vortices, and observe through numerical simulations how $\nabla\times\boldsymbol{\omega}$ mediates the transfer across scales.}

\HLL{
\subsection{Boundary conditions in finite domains} \label{sec:boundarycondfindom}
For problems posed in finite domains, in principle, the following modifications are necessary. Consider the Poisson equation Eq.~\eqref{eq:poissonnobound} subject to a Dirichlet boundary condition on the boundary $\partial \Omega$ of a domain $\Omega$,
\begin{equation}
\overline{\boldsymbol{\omega}}(\mathbf{r}) = h(\mathbf{r}), 
\quad \mathbf{r} \in \partial \Omega.
\end{equation}
In this case, the solution can be expressed using the Green's function 
$G(\mathbf{r}, \mathbf{r}')$ for the domain $\Omega$, which satisfies the boundary condition $\left.G(\mathbf{r}, \mathbf{r}')\right|_{\mathbf{r}\in\partial\Omega}=0$, as
\begin{align}
\overline{\boldsymbol{\omega}}(\mathbf{r})
= -\frac{1}{\eta}
\int_{\Omega} G(\mathbf{r}, \mathbf{r}') 
\, \boldsymbol{S}(\mathbf{r}') 
\, d\mathbf{r}' 
+ \int_{\partial \Omega}
\frac{\partial G(\mathbf{r}, \mathbf{r}')}{\partial n'}
\, h(\mathbf{r}') 
\, dA(\mathbf{r}').
\label{eq:boundarycond}
\end{align}
Here, $\partial/\partial n'$ denotes the normal derivative with respect to $\mathbf{r}'$, and $dA(\mathbf{r}')$ is the surface element on the boundary. This expression shows that the volume integral accounts for the contribution from the source term $\boldsymbol{S}$ within the domain, while the boundary integral captures the contribution from the prescribed Dirichlet boundary value $h$. In this way, even in finite domains with boundary conditions, $\overline{\boldsymbol{\omega}}$ is uniquely determined. 
}

\HLL{The central concern of the present study has been whether the rotational viscosity term mediates information transfer across scales. This is concisely expressed by Eq.~(\ref{eq:TransOmega}) and, more generally, by Eq.~(\ref{eq:poisson_solution}). Equation~(\ref{eq:boundarycond}) provides the necessary theoretical modification for addressing finite-domain problems while maintaining the essential structure of these formulations. Importantly, this modification does not significantly impair the inter-scale transfer property of the rotational viscosity term. Therefore, the theoretical discussions in this section based on the infinite-domain approximation remain fundamentally valid. However, it is crucial to note that the vorticity field obtained in simulations is inevitably affected by finite-domain boundaries, and differs from the behavior expected in unbounded domains.

In the fluid simulations conducted in this study, we impose Dirichlet boundary conditions $\overline{\boldsymbol{\omega}}(\mathbf{r}) = 0$ on the circular domain boundaries. In numerical simulations, strict enforcement of Dirichlet conditions often leads to discontinuous gradients near the boundaries, which can generate spurious numerical oscillations or reflected waves. These artifacts may adversely affect the solution of the Poisson equation or the computed velocity field, potentially compromising convergence and numerical stability.}
\HLL{To mitigate this issue, we smoothly attenuate the physical fields (such as velocity and vorticity) near the domain boundaries. This eliminates discontinuities at the outer edge of the computational domain, thereby improving both numerical stability and solution accuracy. The required smoothness depends on the problem, but in this study---where we focus on rotational viscosity involving high-order spatial structures such as $\nabla \times \boldsymbol{\omega}$---we employ a smootherstep function~\cite{ebert2002texturing} with $C^2$ continuity to smoothly reduce field values to zero near the boundaries. Defined over the domain $t \in [0, 1]$, the function is given by}
\begin{equation}
\HLL{w(t) = 1 - t^3 (10 - 15t + 6t^2).}
\end{equation}
\HLL{This function ensures continuity of the value, first derivative, and second derivative at both endpoints $t = 0$ and $t = 1$, effectively suppressing numerical oscillations and reflections at boundaries. As a result, this attenuation strategy provides a more natural approximation to boundary behavior in the discretized partial differential equations and contributes to the stability of both the Poisson solver and the time integration schemes.}

\section{Rethinking the non-solid rotativity} \label{seq:rethinknonsolid}
In continuum mechanics, the velocity gradient tensor $\nabla \vec{u}$ can be decomposed into the rate-of-strain tensor $\vec{E}$ and the rate-of-rotation tensor $\boldsymbol{\omega}$ as follows:
\begin{equation}
    \nabla \vec{u} = \vec{E} + \boldsymbol{\omega}.
\end{equation}
Here, $\vec{E}$ is a symmetric tensor satisfying $\vec{E}^\mathsf{T} = \vec{E}$, while $\boldsymbol{\omega}$ is an antisymmetric tensor satisfying $\boldsymbol{\omega}^\mathsf{T} = -\boldsymbol{\omega}$. Now, consider a reference point $\vec{x}_0$, the velocity field $\vec{u}$ in the vicinity of $\vec{x}_0$ can be expanded in a Taylor series up to third order, retaining the residual terms, as follows:

\begin{eqnarray}
\vec{u} = \vec{u}_{0} + \nabla \vec{u} \cdot \Delta \vec{x}+ \frac{1}{2} \frac{\partial^2 u_m}{\partial x_j \partial x_k} \Delta x_j \Delta x_k + R^{(3)}_{m}. \label{eq:TaylorNormal}
\end{eqnarray}
For notational convenience, higher-order terms beyond the first order are expressed in tensor form for each velocity component $u_m$ $(m = 1, 2, 3)$. We have introduced Levi--Civita symbol and Einstein's summation convention. We define the position offset as $\Delta \vec{x} := \vec{x} - \vec{x}_{0} = (\Delta x_1, \Delta x_2, \Delta x_3)$, and $R^{(3)}_m$ denotes the third-order residual term associated with the $m$-th component of the velocity. $\nabla \times \vec{u}_{0} = 0$ because $\vec{u}_{0}$ is a constant vector. Substituting $\nabla \vec{u} = \vec{E} + \boldsymbol{\omega}$ into Eq.~(\ref{eq:TaylorNormal}), and applying standard vector and tensor calculus, we find that the contribution from the symmetric tensor $\vec{E}$ vanishes due to its self-cancellation under symmetry, leading to the following:

\begin{eqnarray}
\nabla \times \vec{u} = 2 \boldsymbol{\omega} 
+ \varepsilon_{i n m} \frac{\partial^2 u_m}{\partial x_j \partial x_n} \Delta x_j + \tilde{R}^{(3)}_{m}, \label{eq:TaylorCurl}
\end{eqnarray}
where $\tilde{R}^{(3)}_{m}$ represents the rotation of $R^{(3)}_m$ for $m$-th component of velocity. Here, we used the identity $\nabla \times (\vec{A} \cdot \vec{b}) = -(\vec{A}^\mathsf{T} - \vec{A})$ for tensor $\vec{A}$ and vector $\vec{b}$, where $\vec{A} = \vec{E} + \boldsymbol{\omega}$ and $\vec{b} = \Delta \vec{x}$ in our case. The angular velocity vector $\boldsymbol{\omega}_{0}$ and the antisymmetric tensor $\boldsymbol{\omega}$ are related by equation $\omega^{(0)}_{k} = \epsilon_{ijk} \Omega_{ij}$, where the lower subscript $0$ is shifted to the upper subscript in component expression. In the context of rotational transformations, $\boldsymbol{\omega}$ can be considered a vector of reduced dimensionality, and it is reasonable to equate $\boldsymbol{\omega}$ with the angular velocity vector $\boldsymbol{\omega}_{0}$ ($\boldsymbol{\omega}\equiv \boldsymbol{\omega}_{0}$).

An important point to note here is that, although the third term on the right-hand side of Eq.~(\ref{eq:TaylorNormal}) is proportional to the square of the infinitesimal displacement $\Delta x_a$ ($a = i, j, \text{or } k$), the second term on the right-hand side of Eq.~(\ref{eq:TaylorCurl}), after the curl operation is applied, becomes linearly proportional to $\Delta x_a$. Therefore, the difference between the left-hand side and the first term on the righ-hand side $\boldsymbol{\omega} - 2\boldsymbol{\omega}_0$ is directly proportional to $\Delta x_a$ and is not negligible in most cases. Consequently, it is physically meaningful to incorporate the deviation from solid-body rotation into the antisymmetric part of the stress tensor $\vec{T}_a$, which can be achieved by defining $\vec{T}_a$ as $\vec{T}_a = \eta_r (\boldsymbol{\omega} - 2\boldsymbol{\omega}_0)$, where $\eta_r$ is the rotational viscosity coefficient.

\begin{figure}[t]
\begin{center}
\vspace{-0.2cm}
\includegraphics[width=0.485\textwidth, clip, bb= 0 0 826 779 ]{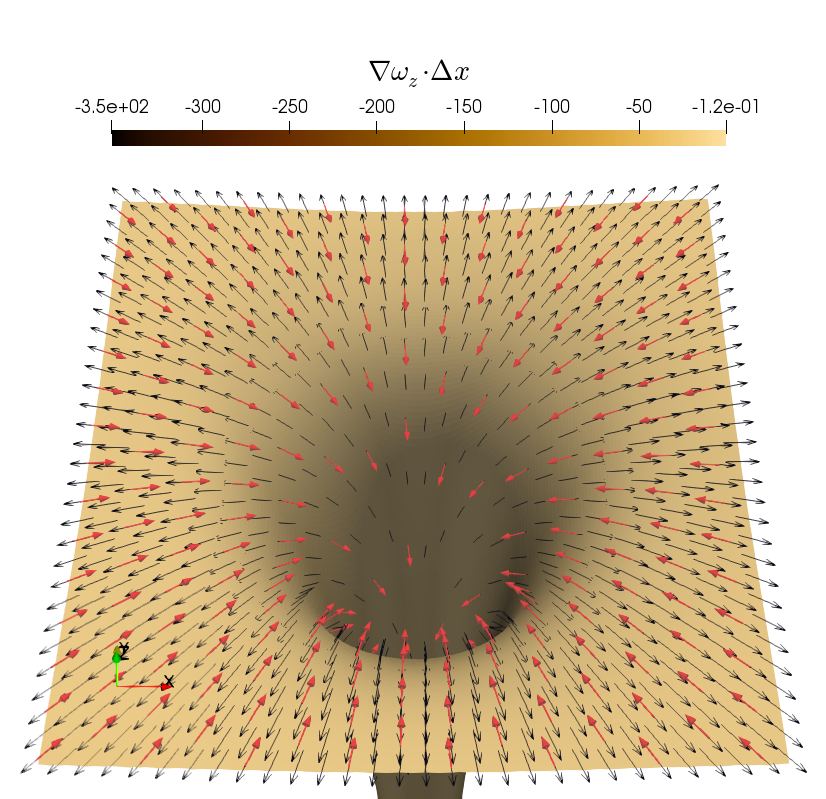}
\end{center}
\caption{A plot of $\nabla \omega_z \cdot \Delta \vec{x}$, visually confirming that $\nabla \omega_z \cdot \Delta \vec{x}$ is always negative for a stable vortex system with the vorticity distribution of $\omega_z(\vec{r}) = 1/r$. The red and gray colored arrows represent the vectors $\nabla \omega_{z}$ and $\Delta \vec{x}$, respectively. }
\label{fig:Figure1}
\end{figure}

Through the rotation operation, the order of dependence on $\Delta x_a$ in each term of Eq.~(\ref{eq:TaylorCurl}) is reduced. Specifically, the second term becomes linearly proportional to the infinitesimal vector $\Delta x_a$, while the residual term $\tilde{R}^{(3)}_m$ depends at least on the square of $\Delta x_a$. Neglecting the residual term, the relation $\boldsymbol{\omega} - 2\boldsymbol{\omega}$ becomes equal to the second term. Therefore, the non-vanishing nature of $\boldsymbol{\omega} - 2\boldsymbol{\omega}$ is characterized by this second term. It can be expressed as the product of a $3 \times 3$ matrix $\vec{C}$, which consists of the second derivatives of the velocity field $\vec{u} = (u_1, u_2, u_3)$, and the infinitesimal displacement vector $\Delta \vec{x}$, such that $\vec{C} \cdot \Delta \vec{x}$. (See Appendix for derivation details.) As an illustrative example where $\vec{C} \cdot \Delta \vec{x} \neq 0$, consider a two-dimensional domain containing a single point vortex. Assume the point vortex is placed at the origin and rotates counterclockwise with constant angular velocity, oriented upward normal to the $xy$-plane. Since this is a two-dimensional case, we have $u_3 = 0$, and the velocity gradients in the $z$-direction vanish for all components.
Under this setting, only the components of $\vec{C}$ involving partial derivatives of $u_1$ and $u_2$ with respect to $x$ or $y$ remain; in particular, only the $C_{31}$ and $C_{32}$ components are nonzero, while all other components of $\vec{C}$ are zero. By the problem setup, the vorticity vector is given as $\boldsymbol{\omega} = (0, 0, \omega_{z})$, where $\omega_{z}$ is assumed to follow an inverse radial profile such as $\omega_{z}(\vec{r}) = 1/r$, with $r = |\vec{r}|$ denoting the distance from the origin.
Since this is a two-dimensional system, we write $\Delta \vec{x} = (\Delta x_1, \Delta x_2)$. After a few algebraic steps, we find
\begin{eqnarray}
\vec{C} \cdot \Delta \vec{x} &=& (\partial_x \omega_z)\Delta x_1 + (\partial_y \omega_z)\Delta x_2 \nonumber \\
		&=& \nabla \omega_z \cdot \Delta \vec{x}. 
\end{eqnarray}
Using $\omega_z = 1/r$, we obtain $\nabla \omega_z = -\vec{r}/|\vec{r}|^3$. Letting $\Delta \vec{x} = \vec{r}$, it follows that $\nabla \omega_z \cdot \Delta \vec{x} = -1/|\vec{r}| < 0$, which holds for all $\vec{r} \neq \vec{0}$. Figure~\ref{fig:Figure1} shows the plot of $\nabla \omega_z \cdot \Delta \vec{x}$, visually confirming that $\nabla \omega_z \cdot \Delta \vec{x}$ is always negative. Recall $\boldsymbol{\omega}\equiv \boldsymbol{\omega}_{0}$. In summary, the expression $\vec{T}_{a} = \eta_{r} (\boldsymbol{\omega} - 2\boldsymbol{\omega}_0$) is always nonzero in a two-dimensional point vortex system that contains only one point vortex.

Equation~(\ref{eq:TransOmega}) gives the vorticity at position $\vec{r}$ as a convolution integral of the contributions from the surrounding subscale vortices; thus, we can evaluate the moments exerted by the surrounding vortices at position $\vec{r}$ by employing a multipole expansion of the right-hand side of Eq.~(\ref{eq:TransOmega}) using the Legendre polynomials $P_n(x)$ as follows.
\begin{eqnarray}
\vec{\overline{\omega}}(\vec{r}) &=& \frac{1}{2C}\sum_{n=0}^{\infty} \frac{1}{r^{n+1}} \int r'^n P_n(\hat{\vec{r}} \cdot \hat{\vec{r}}') \boldsymbol{\omega}(\vec{r}') \, d \vec{r}' \nonumber \\
&=& \frac{1}{2C r} \int \boldsymbol{\omega}(\vec{r}') d\vec{r}' + \nonumber \\
&~& \frac{1}{2C r^3} \int \boldsymbol{\omega}(\vec{r}') (\vec{r}' \cdot \vec{r}) d\vec{r}' + \cdots  
\label{eq:MultiPoleLegendre}
\end{eqnarray}
The first term on the right-hand side represents the monopole term, which corresponds to the total sum of vorticity over the entire domain. In systems exhibiting dipole-like structures---where positive and negative vortices are paired---this term cancels out. In contrast, when all vortices within the domain rotate in the same direction, the contributions do not cancel, and the monopole term remains nonzero. 
The second term is the dipole term, which represents the first raw moment of vorticity over the domain. It characterizes the influence of spatial bias in the vorticity distribution at position $\vec{r}$. For example, in a centrosymmetric configuration---where the vorticity is balanced in both horizontal and vertical directions---this term cancels and vanishes. Otherwise, it contributes a finite effect. Higher-order terms (from the third onward) represent finer features of the vorticity distribution, such as geometric deformation or biaxial asymmetry. 

Taken together, $\boldsymbol{\omega} - 2\boldsymbol{\omega}_0$ is linearly proportional to the infinitesimal displacement vector and, therefore, is not negligible. In addition, $\boldsymbol{\omega} - 2\boldsymbol{\omega}_0$ can be nonzero in the vicinity of point vortices. The non-vanishing property of $\boldsymbol{\omega} - 2\boldsymbol{\omega}_0$, that is,
$\exists\, \boldsymbol{\omega} - 2\boldsymbol{\omega}_0 \in \mathbb{R},\quad \boldsymbol{\omega} - 2\boldsymbol{\omega}_0 \neq 0$,
is one of the two necessary conditions required to ensure that Eq.~(\ref{eq:NSwithSAMC}) has physical significance. The other condition is the existence of a nonzero rotational viscosity:
$\exists\, \nabla \times \boldsymbol{\omega} \in \mathbb{R},\quad \nabla \times \boldsymbol{\omega} \neq 0$.
Since interscale transfer is mediated by rotational viscosity $\nabla \times \boldsymbol{\omega}$, Eq.~(\ref{eq:RecurrRel}) would lose validity if the magnitude of $\nabla \times \boldsymbol{\omega}$ were identically zero.
We take the rotation of Eq.~(\ref{eq:MultiPoleLegendre}) and examine the expansion of $\nabla \times \boldsymbol{\omega}$. A straightforward calculation reveals that the first-order term of $\nabla \times \boldsymbol{\omega}$ is given by
\begin{align}
\nabla \times \boldsymbol{\omega}^{(1)} = -\frac{\vec{r}}{r^3} \times \vec{\Gamma},
\end{align}
where $\vec{\Gamma}$ is the integrated vorticity over the domain and is equal to the first term on the right-hand side of Eq.~(\ref{eq:MultiPoleLegendre}). Mathematically, for $\nabla \times \boldsymbol{\omega}^{(1)}$ to be nonzero at position $\vec{r}~(\ne \vec{0})$, two conditions must be met: (i) the vectors $\vec{r}$ and $\vec{\Gamma}$ must not be parallel, and (ii) the magnitude of $\vec{\Gamma}$ must not vanish.
In 2D systems, the vorticity vector always points perpendicular to the $xy$-plane, so condition (i) is trivially satisfied. On the other hand, condition (ii) is satisfied in systems composed entirely of positive-signed vortices, as mentioned above. Thus, in such 2D systems, $\nabla \times \boldsymbol{\omega}^{(1)}$ is guaranteed to be nonzero.

In real systems, where point vortex models apply or in cases such as quantum vortices observed in a phase-controlled state---where vortex lines extend in the out-of-plane direction and rotate coherently---the nonzero nature of the first-order term can be expected to hold. In contrast, higher-order terms arise from asymmetries and deformations in vorticity distribution and may enhance or suppress the contribution from the first-order term, depending on the behavior of these higher-order structures.

Equation~(\ref{eq:RecurrRel}) illustrates that the transfer of vorticity across scales can occur recursively through the rotational viscosity term $\nabla \times \boldsymbol{\omega}$. The theoretical analysis in this section reveals that three conditions must be satisfied for this effect to emerge. First, the vorticity gradient ($\nabla \boldsymbol{\omega}$) must become sufficiently steep to yield a nonzero value of $\boldsymbol{\omega} - 2\boldsymbol{\omega}_0$, potentially resulting in non-solid-body rotational flows. Second, one of the following subconditions must hold: (i) the total vorticity over the domain is nonzero, and the higher-order moments of the vorticity distribution do not cancel its contribution; or (ii) the total vorticity is zero, while the moments of the vorticity distribution remain nonzero. Third, the attenuation effect introduced by filtering, represented by ${\mathcal F}$ (or ${\mathcal F} \cdot {\mathcal G}$), must be sufficiently small so as not to negate the preceding conditions. Viscous dissipation is expected to be the primary contributor to this attenuation. We have also shown that a system composed entirely of positively signed point vortices can simultaneously satisfy all three of these conditions. 

Moreover, the interscale transfer of vorticity can also be observed within the framework of the standard Navier--Stokes formulation. Equation~(\ref{eq:NSwithSAMC}) is mathematically equivalent to the conventional Navier--Stokes equations; its vorticity-explicit form does not alter the underlying dynamics. Therefore, in the following section, we investigate a freely decaying 2D turbulent flow in a finite domain initialized with a configuration of positive-signed point vortices. This analysis is conducted using the vorticity--streamfunction formulation of the Navier--Stokes equations for incompressible viscous fluids, in order to further explore the nature of the interscale transfer mechanism.

\section{Simulation of freely decaying turbulence} \label{sec:simulation}
{\it Problem setup.}---We consider 2D vorticity transport in an incompressible viscous fluid confined to a circular domain with the diameter of $L$. The flow field is described using the vorticity--streamfunction formulation:
\begin{equation}
\HLL{\frac{\partial \boldsymbol{\omega}}{\partial t} + (\mathbf{u} \cdot \nabla) \boldsymbol{\omega} = \eta \nabla^2 \boldsymbol{\omega}}.
\label{eq:time_evolution}
\end{equation}
As a preparatory step, the initial point vortex distribution used in the vorticity transport simulation is produced as follows. First, $N=100$ positively signed point vortices, all with identical orientation, are randomly placed within a circular region of radius $R = L/2$ in the domain. Each vortex is assigned a nondimensional circulation strength of $\Gamma = 1$. To enforce the boundary condition on the circular edge, the method of image vortices is employed. We then computed the evolution of the vortices by calculating the interaction between each pair of vortices using the Biot-Savart law, and then integrated the results over time using the fifth-order Runge--Kutta--Fehlberg method~\cite{Fehlberg1969}. All vortex-vortex interactions are directly calculated without approximation. We verify that the relative error in the total interaction energy remains within approximately 0.63\% throughout the simulation. 

\HLL{The resulting velocity field is then used as the initial condition for computing the vorticity--streamfunction formulation. In this computation, a two-dimensional Cartesian grid is defined over a rectangular domain, while a Dirichlet boundary condition $\boldsymbol{\omega} = 0$ is imposed along the circular boundary of radius $R$ throughout the simulation. In addition, the exterior region outside the circular boundary---i.e., the unused margin of the rectangular domain---is also filled with $\boldsymbol{\omega} = 0$. As a result, the effective computational domain for the vorticity--streamfunction formulation remains circular in shape. To ensure numerical stability, a damping procedure is applied near the circular boundary so that the vorticity distribution smoothly decays as a function of distance from the boundary. This attenuation prevents discontinuities and spurious gradients at the edge and promotes stable convergence of the numerical solution. For further details on the specific boundary treatment, see Section~\ref{sec:boundarycondfindom}.} The parameter $R$ is set to 1, which directly yields $L = 2$. All lengths are thus nondimensionalized with respect to $R$. In this setup, the nondimensionalization is performed using $R$ for length and $\Gamma$ for circulation. Consequently, time and kinematic viscosity are nondimensionalized as
\begin{align}
\tilde{t} = \frac{\Gamma}{2\pi R^2} t, \quad \tilde{\nu} = \frac{2\pi}{\Gamma} \nu,
\end{align}
where $t$ and $\nu$ denote the dimensional time and kinematic viscosity, respectively.

{\it Numerical conditions.}---We conduct simulations of freely decaying 2D turbulence in a finite domain using three different values of the nondimensional kinematic viscosity: $\tilde{\nu} = 1.0 \times 10^{-2}$, $1.0 \times 10^{-3}$, and $1.0 \times 10^{-4}$. The square domain is discretized using a uniform Cartesian grid with resolution $(n_x, n_y) = (2048, 2048)$, where $n_x$ and $n_y$ denote the number of grid points in each direction. 
The vorticity equation for incompressible flow is discretized in space using second-order central finite differences. For time integration, the implicit Crank--Nicolson scheme~\cite{Crank_Nicolson_1947, doi:10.1137/0727022} is adopted to achieve second-order temporal accuracy. The nonlinear advection term is evaluated at each time step using fixed-point iteration (Picard iteration).
The velocity field is recovered as the perpendicular gradient of the streamfunction. The incompressibility constraint is enforced by solving the Poisson equation for the streamfunction (computed from vorticity) using the successive over-relaxation (SOR) method. The Red-Black algorithm~\cite{adams1982multi} is implemented to ensure compatibility with parallel computation. The time step is initially set to $\Delta \tilde{t} = 1.0 \times 10^{-4}$ and adaptively adjusted to satisfy the Courant--Friedrichs--Lewy (CFL) condition based on the maximum velocity at each time step. A total of 6000 time steps were computed. Assuming physical parameters of circulation $\Gamma = 1.0 \times 10^{-3}$~cm$^2$/s and radius $R = 1.0$~cm, this corresponds to approximately 3770 seconds (about 1 hour and 3 minutes) of physical time under freely decaying conditions.

\begin{figure*}[t]
\begin{center}
\vspace{-0.5cm}
\centering
\includegraphics[width=1.08\textwidth, clip, bb= 0 0 790 550 ]{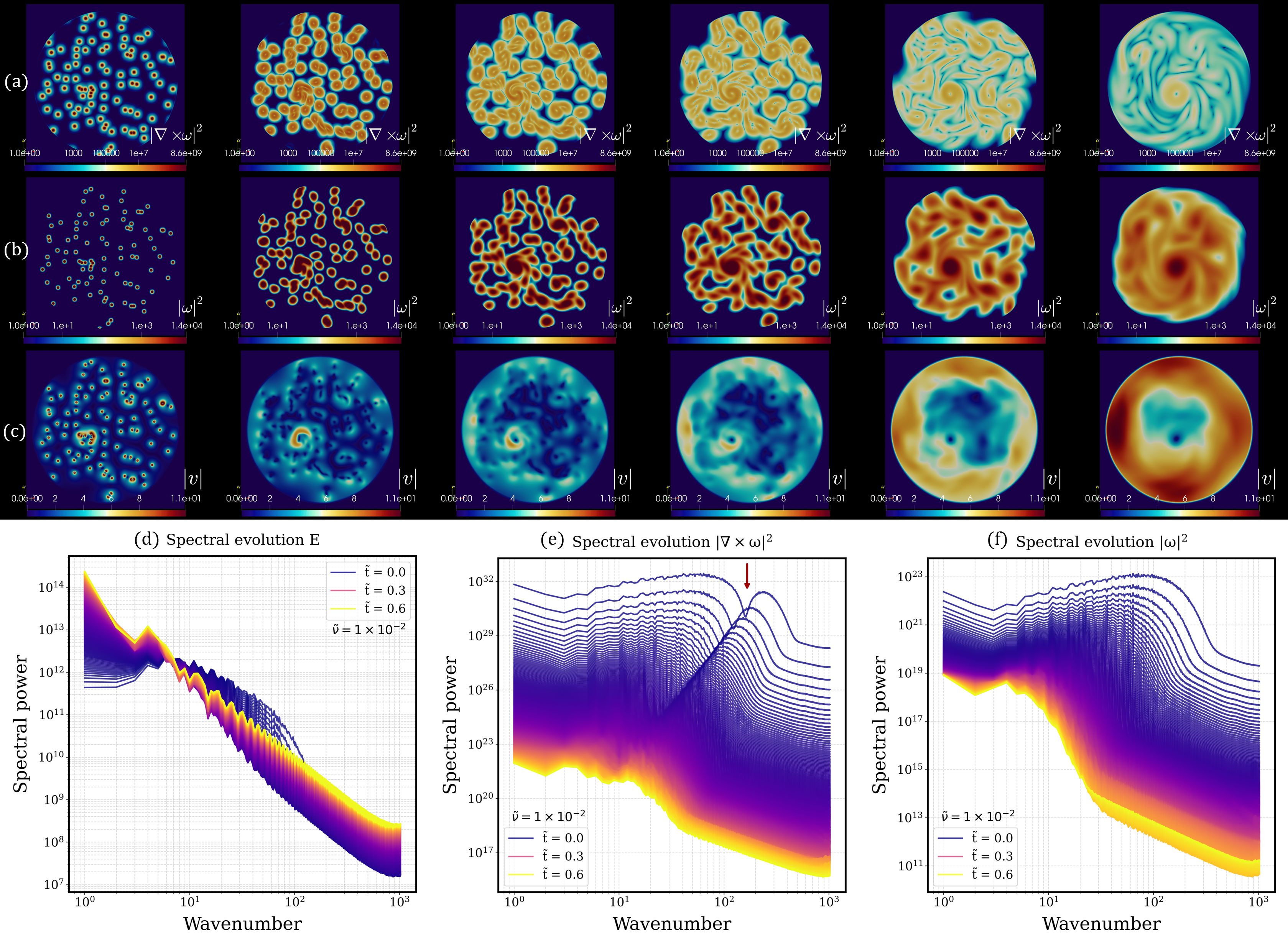}
\caption{\HLL{Simulation results for the high-viscosity case \(\tilde{\nu} = 1.0 \times 10^{-2}\). Panels~(a), (b), and~(c) show snapshots of the spatial distribution of the rotational vorticity magnitude \(|\nabla \times \boldsymbol{\omega}|^2\), enstrophy \(|\boldsymbol{\omega}|^2\), and velocity norm \(|\boldsymbol{u}|\) (denoted as \(|v|\)), respectively, taken within the time interval \(\tilde{t} \leq 0.6\). Panels~(d), (e), and~(f) depict the corresponding spectral evolution of kinetic energy, rotational vorticity magnitude, and enstrophy, respectively, from dimensionless time \(\tilde{t} = 0.0\) to \(\tilde{t} = 0.6\).}}
\label{fig:Figure2}
\end{center}
\end{figure*}

\begin{figure*}[t]
\begin{center}
\vspace{-0.5cm}
\centering
\includegraphics[width=1.08\textwidth, clip, bb= 0 0 790 550 ]{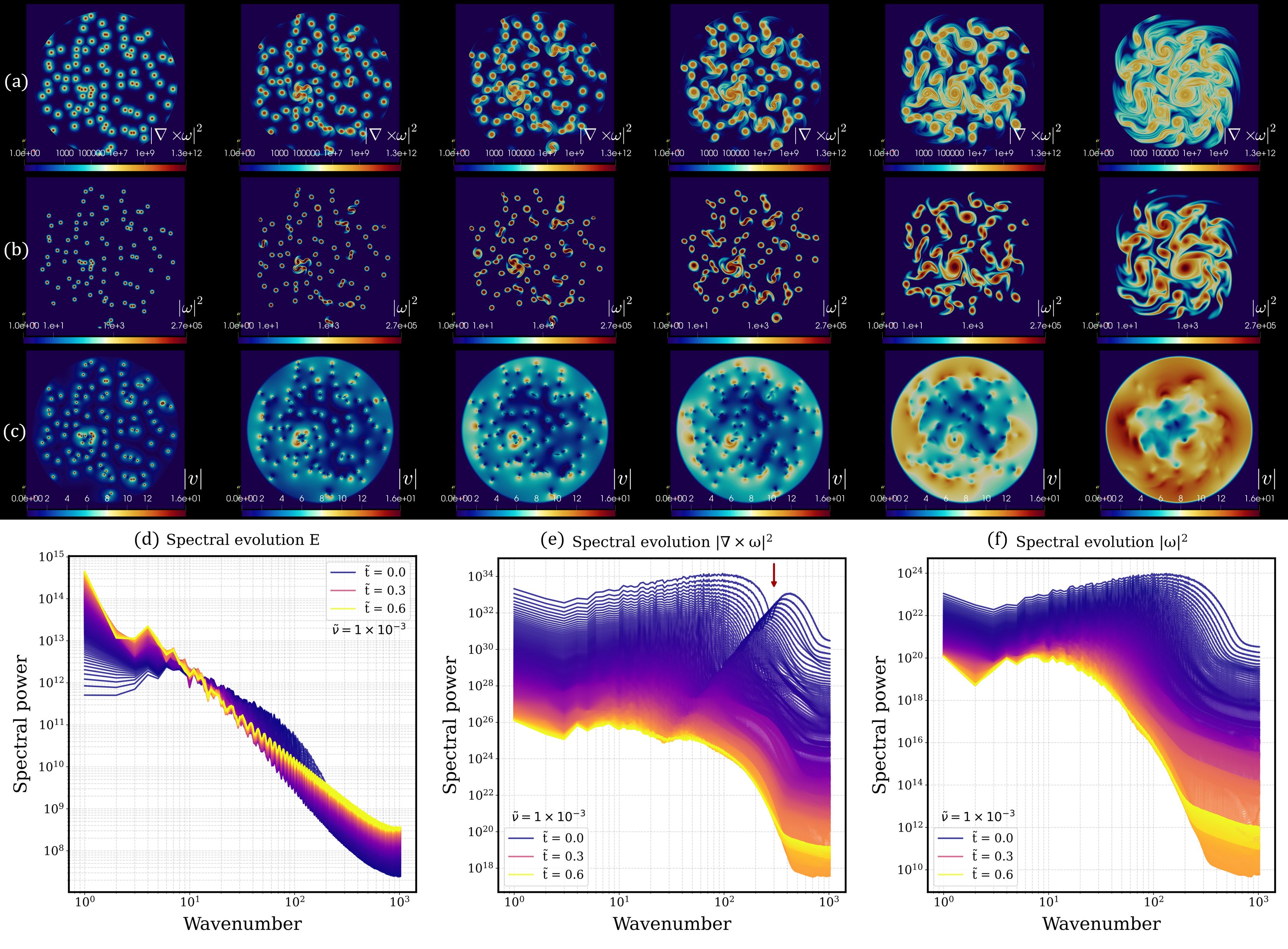}
\caption{\HLL{Simulation results for the intermediate-viscosity case \(\tilde{\nu} = 1.0 \times 10^{-3}\). Panels~(a), (b), and~(c) present the spatial fields of \(|\nabla \times \boldsymbol{\omega}|^2\), 
\(|\boldsymbol{\omega}|^2\), and \(|\boldsymbol{u}|\), respectively. Panels~(d)--(f) show the spectral evolution of energy, rotational vorticity magnitude, and enstrophy over the simulation period \(\tilde{t} = 0.0\) to \(\tilde{t} = 0.6\).}}
\label{fig:Figure3}
\end{center}
\end{figure*}

\begin{figure*}[t]
\begin{center}
\vspace{-0.5cm}
\centering
\includegraphics[width=1.08\textwidth, clip, bb= 0 0 790 550 ]{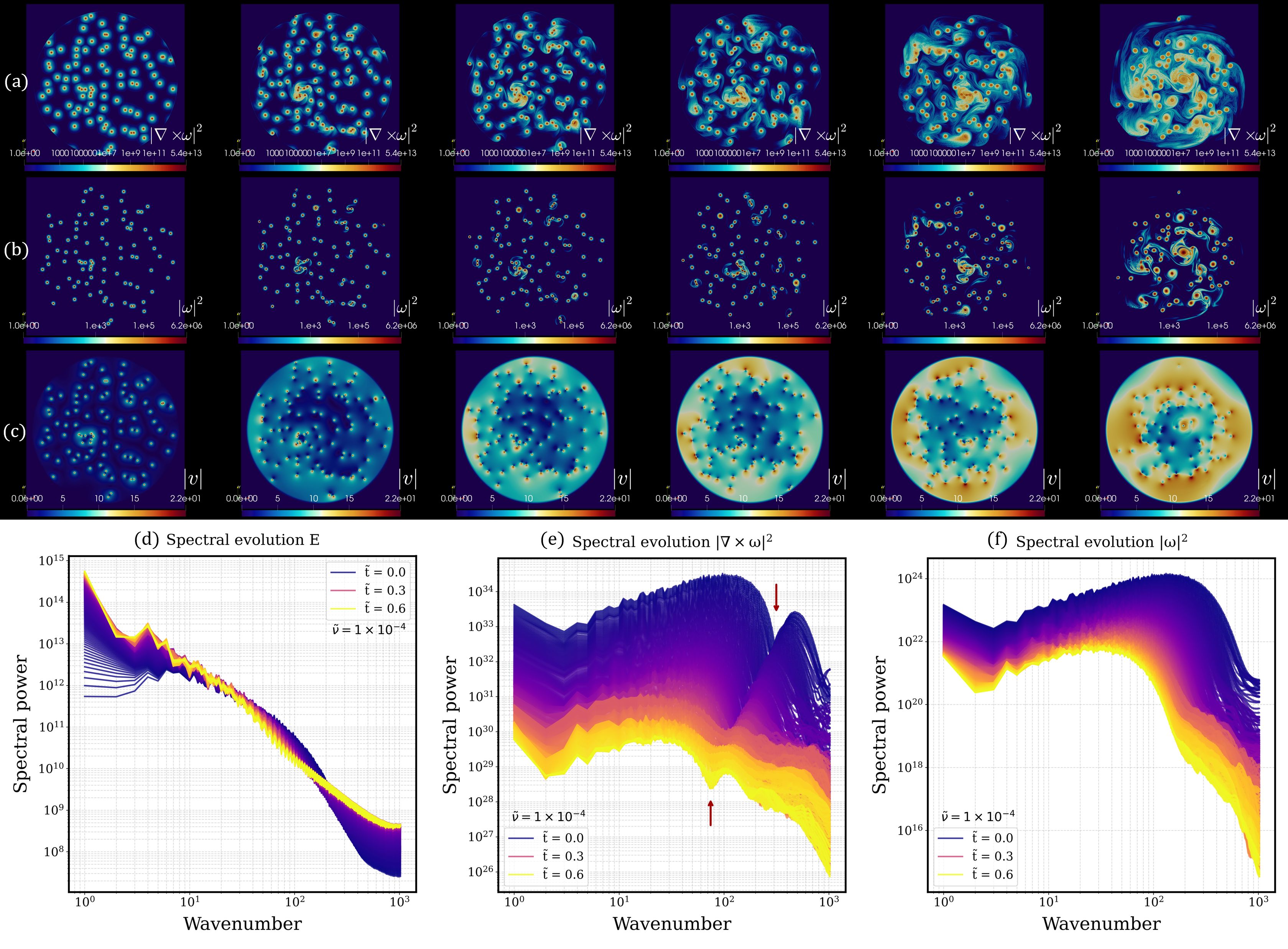}
\caption{\HLL{Simulation results for the low-viscosity case \(\tilde{\nu} = 1.0 \times 10^{-4}\). Panels~(a)--(c) visualize the distributions of \(|\nabla \times \boldsymbol{\omega}|^2\), \(|\boldsymbol{\omega}|^2\), and \(|\boldsymbol{u}|\), respectively. The time evolution of their corresponding spectra is displayed in panels~(d)--(f), showing the development from \(\tilde{t} = 0.0\) to \(\tilde{t} = 0.6\).}}
\label{fig:Figure4}
\end{center}
\end{figure*}

\HLL{{\it Results.}---Figures~\ref{fig:Figure2}--\ref{fig:Figure4} present the simulation results for three different viscosity conditions: (i)~$\tilde{\nu} = 1.0 \times 10^{-2}$, (ii)~$1.0 \times 10^{-3}$, and (iii)~$1.0 \times 10^{-4}$. Note that, although the computational domain is circular, as described above, Figs.~2--4 display the entire two-dimensional Cartesian grid. In each figure, panels~(a), (b), and~(c) show the snapshots of the magnitude of rotational viscosity $|\nabla \times \boldsymbol{\omega}|^2$, the vorticity magnitude (enstrophy) $|\boldsymbol{\omega}|^2$, and the velocity norm $|\boldsymbol{u}|$ (denoted as $|v|$ in the figures), respectively. Panels~(d), (e), and~(f) in each figure illustrate the spectral evolution from the beginning of the simulation up to $\tilde{t} = 0.6$, for the kinetic energy, the rotational vorticity strength $|\nabla \times \boldsymbol{\omega}|^2$, and the enstrophy $|\boldsymbol{\omega}|^2$, respectively. These visual and spectral data collectively provide a comprehensive view of how viscosity governs the multiscale structure and energy dynamics of decaying 2D turbulence.

A clear hierarchy in spatial complexity is observed across the panels (a)--(c) of each figure. Specifically, $|\nabla \times \boldsymbol{\omega}|^2$ in panel (a) displays the most intricate spatial distribution, followed by $|\boldsymbol{\omega}|^2$ in panel (b), and finally $|\mathbf{v}|$ in panel (c), which exhibits the smoothest and broadest structures. This progression is consistent with the mathematical nature of the quantities: $|\mathbf{v}|$ represents the base velocity field, while $|\boldsymbol{\omega}|^2$ corresponds to its first-order spatial derivative (vorticity), and $|\nabla \times \boldsymbol{\omega}|^2$ involves second-order derivatives of velocity. Hence, higher-order spatial derivatives naturally encode more localized and rapidly varying features, making panel (a) the most sensitive to small-scale structures. The contrast across these fields provides direct evidence that $|\nabla \times \boldsymbol{\omega}|^2$ captures the highest degree of spatial complexity, underscoring its value in visualizing fine-scale turbulent dynamics and dissipation mechanisms.

In Fig.~\ref{fig:Figure2}, where the viscosity is relatively high ($\tilde{\nu} = 1.0 \times 10^{-2}$), the spatial plots (a--c) show a rapid decay of fine-scale structures. The $|\nabla \times \boldsymbol{\omega}|^2$ field quickly becomes smooth and diffuse, indicating that small-scale rotational structures are efficiently eliminated by viscous dissipation. The enstrophy field ($|\boldsymbol{\omega}|^2$) follows a similar trend, revealing the merging and weakening of vortex cores over time. Correspondingly, the velocity norm shows a gradual emergence of large-scale coherent flow patterns as small-scale fluctuations are suppressed. The spectral evolution plots (d--f) clearly support these observations: spectral energy and enstrophy at high wavenumbers decay steeply, and the spectral slope becomes steeper with time. Notably, the spectral peak of $|\nabla \times \boldsymbol{\omega}|^2$ gradually shifts toward lower wavenumbers over time, reflecting the suppression of small-scale fluctuations and a transition toward larger-scale structures under strong viscous damping. In this regime, viscosity dominates the dynamics, suppressing nonlinear interactions and enforcing a monotonic and irreversible decay of energy.

As the viscosity is reduced to $\tilde{\nu} = 1.0 \times 10^{-3}$ in Fig.~\ref{fig:Figure3}, the persistence of small-scale structures becomes apparent. In the spatial plots, the $|\nabla \times \boldsymbol{\omega}|^2$ field retains pronounced filamentary features, while the enstrophy field shows ongoing vortex merging and localized concentrations. The velocity field also develops a richer structure with intermediate-scale eddies and spiral-like formations. The corresponding spectra reveal critical differences: in particular, the $|\nabla \times \boldsymbol{\omega}|^2$ spectrum (Fig.~\ref{fig:Figure3}e) exhibits a clear peak that migrates from high to lower wavenumbers over time. This drift is indicative of the system's capacity to redistribute rotational intensity from fine to coarser spatial scales. Rather than merely dissipating, small-scale vortical structures are seen to undergo nonlinear interactions---merging, stretching, and reorienting---resulting in the growth of larger-scale coherent structures. The spectral peak movement can be interpreted as a manifestation of the inverse energy cascade, a hallmark of 2D turbulence in which energy flows from small to large scales even as enstrophy continues to dissipate toward smaller scales. This regime exemplifies a balance between viscous damping and nonlinear self-organization.

In the low-viscosity case of Fig.~\ref{fig:Figure4} ($\tilde{\nu} = 1.0 \times 10^{-4}$), these dynamics become more pronounced and complex. Spatially, the $|\nabla \times \boldsymbol{\omega}|^2$ field features densely packed and intricate filamentary patterns that persist well into later times. Vortices retain their identity longer and exhibit dynamic interactions including fusion and alignment, visible in the enstrophy field. The velocity norm likewise reveals a highly intricate flow structure with clear signs of multiscale interactions. Spectrally, the system exhibits strong signatures of nonlinear turbulence. The $|\nabla \times \boldsymbol{\omega}|^2$ spectrum (Fig.~\ref{fig:Figure4}e) displays multiple peaks and non-monotonic features, with primary peaks shifting toward lower wavenumbers while maintaining high-frequency power. This suggests a two-tiered mechanism: while some energy is transferred upscale in the form of larger coherent structures, a residual component remains active at smaller scales, possibly due to intermittent vortex filamentation or localized shear layers. This observation provides strong evidence of an active inverse cascade regime coexisting with a weak direct enstrophy cascade, both supported by a minimal level of viscosity. The intricate spectral behavior in Fig.~\ref{fig:Figure4} illustrates the richness and complexity of turbulence in the low-viscosity limit, driven by robust nonlinear interactions across a broad range of scales.

These findings underline the critical role viscosity plays in shaping the structural and spectral evolution of two-dimensional turbulence. At high viscosities, the system quickly relaxes into a smooth, large-scale state, dominated by dissipation. As viscosity decreases, nonlinear dynamics become more influential, enabling small-scale interactions and vortex mergers that lead to the development of coherent large-scale flow. The shifting peak in the $|\nabla \times \boldsymbol{\omega}|^2$ spectrum from high to low wavenumbers over time does not imply a physical ``movement'' of structures in space but rather a redistribution of rotational intensity in scale space---an effective scale reorganization. Particularly in the low-viscosity regime, the ability of the system to sustain multiscale coherence and transfer rotational energy across scales becomes dominant, aligning with classical theoretical predictions of the inverse cascade in 2D turbulence. The combination of spatial and spectral diagnostics in Figs.~\ref{fig:Figure2}--\ref{fig:Figure4} thus offers a detailed and compelling view into the mechanisms by which dissipation and nonlinearity jointly govern turbulence decay, highlighting how even a seemingly simple system like 2D vortex dynamics can give rise to remarkably rich and scale-dependent behaviors.}

\begin{figure*}[t]
\begin{center}
\vspace{-0.5cm}
\centering
\includegraphics[width=1.08\textwidth, clip, bb= 0 0 790 550 ]{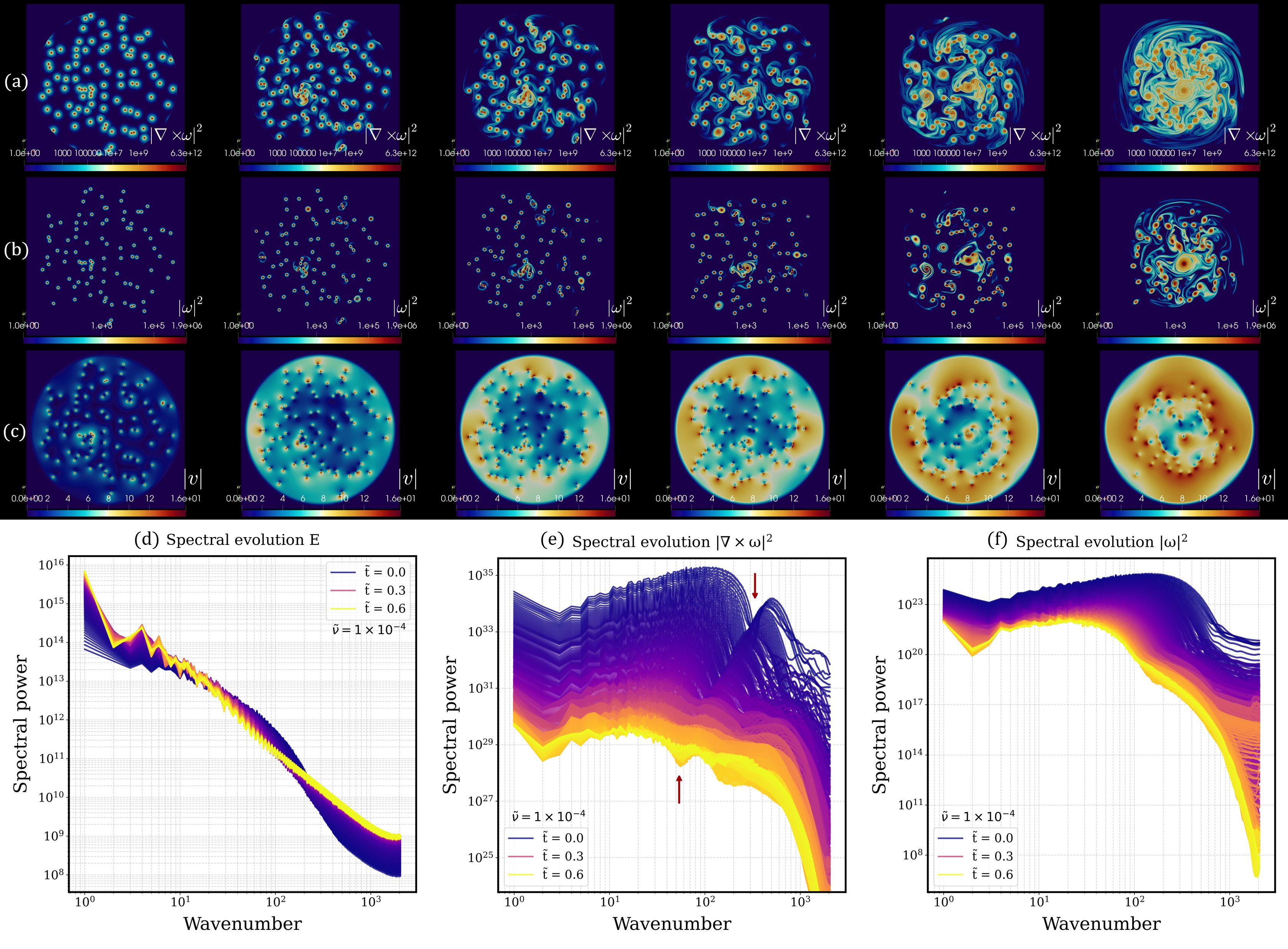}
\caption{\HLL{Simulation results at high resolution (\(4096 \times 4096\)) for the low-viscosity case \(\tilde{\nu} = 1.0 \times 10^{-4}\). Panels~(a), (b), and~(c) show the spatial distributions of the rotational vorticity magnitude \(|\nabla \times \boldsymbol{\omega}|^2\), enstrophy \(|\boldsymbol{\omega}|^2\), and velocity norm \(|\boldsymbol{u}|\) (denoted as \(|v|\)), respectively. Panels~(d), (e), and~(f) display the spectral evolution of kinetic energy, rotational vorticity magnitude, and enstrophy from \(\tilde{t} = 0.0\) to \(\tilde{t} = 0.6\). Compared to lower-resolution cases, finer small-scale features and sharper spectral structures are resolved, indicating enhanced scale interactions and a broader inertial range.}}
\label{fig:Figure5}
\end{center}
\end{figure*}

\HLL{{\it Effects of grid resolution and future directions.}---Figure~\ref{fig:Figure5} presents a high-resolution simulation (\(4096 \times 4096\)) under the same low-viscosity condition \(\tilde{\nu} = 1.0 \times 10^{-4}\) as in Fig.~\ref{fig:Figure4}. While key dynamical features remain consistent with the \(2048 \times 2048\) case, the increased resolution reveals new small-scale structures and spectral behaviors that refine our understanding of scale interactions.
Several trends persist across both resolutions. The spatial fields continue to exhibit a clear hierarchy of complexity: \(|\nabla \times \boldsymbol{\omega}|^2\) shows the finest features, followed by \(|\boldsymbol{\omega}|^2\), and finally the smoother velocity norm \(|\boldsymbol{u}|\). Spectral evolution plots again display high-wavenumber persistence and downward migration of spectral peaks in \(|\nabla \times \boldsymbol{\omega}|^2\), indicating sustained nonlinear activity and inverse cascade behavior. These consistencies confirm the qualitative robustness of the simulation across grid resolutions.
At higher resolution, however, finer filamentary structures emerge in panel~(a), and the spectrum of \(|\nabla \times \boldsymbol{\omega}|^2\) in panel~(e) shows more detailed oscillations and secondary peaks at high wavenumbers. These reflect enhanced representation of the direct enstrophy cascade and suggest a broader inertial range than previously resolved. The sharper spectral features also indicate that some small-scale statistics may not have been fully converged at lower resolution.
These findings highlight the value of high-resolution simulations in capturing both large- and small-scale dynamics more accurately. While the overall energy transfer mechanisms are preserved at \(2048^2\), resolution-dependent differences in dissipation behavior and spectral sharpness suggest the need for systematic convergence studies. Future work should evaluate resolution effects on higher-order moments, scaling laws, and dissipation structures, and consider adaptive or multiscale approaches for computational efficiency in low-viscosity regimes.}

\HLL{These findings collectively indicate that rotational viscosity, through the behavior of $\nabla \times \boldsymbol{\omega}$, plays an active role in turbulent scale dynamics. In summary, the numerical simulations offer novel insights into the dynamical role of rotational viscosity---specifically, how $\nabla \times \boldsymbol{\omega}$ governs inter-scale energy transfer in freely decaying 2D turbulence. Our simulations reveal that the quantity $|\nabla \times \boldsymbol{\omega}|^{2}$ exhibits a consistent downward spectral shift, which corresponds to an inverse transfer mechanism distinct from classical energy cascade processes. This behavior, observed under varying viscosities and validated by high-resolution simulations (see Fig.~\ref{fig:Figure5}), suggests a previously underappreciated role of rotational viscosity in turbulence self-organization. The results provide a new physical interpretation, showing that angular momentum-related transport phenomena can have measurable, emergent effects even in passive fluid systems without invoking active or chiral forcing.}


\section{Conclusion} \label{sec:conclusion}
Rotational viscosity ($\nabla \times \boldsymbol{\omega}$) is not typically emphasized in turbulence studies, which often focus on conserved quantities such as energy and enstrophy. However, we have shown that initializing the system with a point vortex distribution featuring locally concentrated circulation leads to a rotational viscosity spectrum exhibiting a distinct dip in a specific wavenumber range. This structure persists over time while shifting toward lower frequencies, suggesting that small-scale ordered structures are transmitted to larger scales via the rotational viscosity term. To the best of the author's knowledge, this represents the first numerical observation of such behavior. Previous theoretical work predicted recursive interscale transfer of rotational viscosity within the spin angular momentum-conserving Navier--Stokes model, but lacked numerical verification. In addition, the underlying assumption---the existence of local non-solid rotational flows---had not been theoretically justified.

This study addressed both issues. The model was reinterpreted within the turbulence hierarchy framework: small-scale vortices may appear non-solid from a local perspective, yet still appear solid from a large-scale viewpoint. Based on continuum mechanics and vector analysis, we showed that the deviation $\boldsymbol{\omega} - 2\boldsymbol{\omega}_0$ is linearly proportional to first-order spatial variation, and thus non-negligible. These findings support the theoretical validity of the spin-conserving Navier--Stokes equations and suggest that, under certain conditions, small-scale order can be transmitted to large scales via $\nabla \times \boldsymbol{\omega}$. \HLL{Spectral analysis revealed robust upscale transfer of rotational viscosity $\nabla \times \boldsymbol{\omega}$ across all viscosity regimes, with finer small-scale structures and enstrophy distributions captured more clearly at lower viscosities and higher resolutions.}

\appendix
\section{Derivation of the second term in Equation~(\ref{eq:TaylorCurl})}
The $i$-th component of the curl of vector $\vec{u}$ is expressed using the Levi-Civita symbol as follows:
\begin{eqnarray}
(\nabla \times \mathbf{u})_i &=& \varepsilon_{i n m} \frac{\partial u_m}{\partial x_n}, \label{eq:SecOrderDerivDefCurl}
\end{eqnarray}
where $u_{m}$ represents the $m$th component of velocity $\mathbf{u}$. Denote the $k$th-order term in the Taylor series of $u_{m}$ as $u^{(k)}_{m}$~($k=1,2,3$). The third term on the right-hand side of Eq.~(\ref{eq:TaylorNormal}) can be rewritten using the right-hand side of Eq.~(\ref{eq:SecOrderDerivDefCurl}) as follows:
\begin{eqnarray}
(\nabla \times \mathbf{u})^{(2)}_i &=& \varepsilon_{i n m} \frac{\partial u^{(2)}_m}{\partial x_n} \nonumber \\
&=& \varepsilon_{i n m} \frac{\partial}{\partial x_n}\left[ \frac{1}{2} \frac{\partial^2 u_m}{\partial x_j \partial x_k} \Delta x_j \Delta x_k \right] \nonumber \\
&=& \varepsilon_{i n m} \frac{1}{2} \frac{\partial^2 u_m}{\partial x_j \partial x_k} [ \delta_{nj} \Delta x_k + \Delta x_j \delta_{nk} ] \nonumber \\
&=& \varepsilon_{i n m} \frac{1}{2} \left[ \frac{\partial^2 u_m}{\partial x_n \partial x_k} \Delta x_k + \frac{\partial^2 u_m}{\partial x_j \partial x_n} \Delta x_j \right] \nonumber \\
&=& \varepsilon_{i n m} \frac{\partial^2 u_m}{\partial x_j \partial x_n} \Delta x_j \nonumber \\
&=& C_{ij} \Delta x_j. \label{eq:DerivCurlProcs} 
\end{eqnarray}
Here, interchange of partial derivatives is allowed due to the assumption of sufficient smoothness of the velocity field. $C_{ij}$ is a $3 \times 3$ matrix composed of the second-order derivatives of velocity, expressed as follows:
\begin{equation}
\mathbf{C} \coloneqq
\begin{bmatrix}
\partial_{12} u_3 - \partial_{13} u_2 &
\partial_{22} u_3 - \partial_{23} u_2 &
\partial_{32} u_3 - \partial_{33} u_2 \\
\partial_{13} u_1 - \partial_{11} u_3 &
\partial_{23} u_1 - \partial_{21} u_3 &
\partial_{33} u_1 - \partial_{31} u_3 \\
\partial_{11} u_2 - \partial_{12} u_1 &
\partial_{21} u_2 - \partial_{22} u_1 &
\partial_{31} u_2 - \partial_{32} u_1
\end{bmatrix} 
\end{equation}

\section*{Acknowledgment}
This study was supported by JSPS KAKENHI Grant Number 22K14177 and JST PRESTO, Grant Number JPMJPR23O7. The author acknowledges the use of ChatGPT (developed by OpenAI) and DeepL for initial language writing assistance during manuscript preparation. All AI-assisted content was reviewed and revised by the author. The author would like to thank Editage (www.editage.jp) for English language editing of the final version of the manuscript. The author would like to express sincere gratitude to his family members for their unwavering moral support and encouragement.

\section*{Data availability statement}
No new data were created or analyzed in this study.

\bibliography{main}

\end{document}